%% file: Intensity.tex
\documentclass{tcibook}
\usepackage{fancyhea}
\usepackage{work}
\usepackage{amsmath,amssymb}
\usepackage{psfrag}
\usepackage{threeparttable}
\usepackage{epsfig}  
\usepackage{color}
\usepackage{graphicx}               
\usepackage{url}
\usepackage[normalem]{ulem}
\usepackage{hyperref}
\input workshopsymbols.tex

\usepackage{epstopdf}
\usepackage{multirow}
\usepackage{cite} 

\newcommand{\boss}[2]{\ensuremath{\rlap{\kern-2.5pt\ensuremath{\overset{\scriptscriptstyle(-)}{\phantom{#1}}}}{\ensuremath{{#1}_{#2}}}}}

\begin{document}

\def\bibname{References}
\bibliographystyle{unsrt}

\raggedbottom
\pagenumbering{roman}
\parindent=0pt
\parskip=8pt
\setlength{\evensidemargin}{0pt}
\setlength{\oddsidemargin}{0pt}
\setlength{\marginparsep}{0.0in}
\setlength{\marginparwidth}{0.0in}
\marginparpush=0pt


\pagenumbering{arabic}

\renewcommand{\chapname}{chap:intro_}
\renewcommand{\chapterdir}{.}
\renewcommand{\arraystretch}{1.25}
\addtolength{\arraycolsep}{-3pt}

\thispagestyle{empty}
\begin{centering}
\vfill

{\Huge\bf Planning the Future of U.S. Particle Physics}

{\Large \bf Report of the 2013 Community Summer Study}

\vfill

{\Huge \bf Chapter 2:  Intensity Frontier}

\vspace*{2.0cm}
{\Large \bf Conveners:  J. L. Hewett and H.~Weerts}

\vfill

\input mainauthorlist.tex

\vfill

\end{centering}

\pagenumbering{roman}

\newpage
\mbox{\null}

\vspace{3.0cm}

{\Large \bf Authors of Chapter 2:}

\vspace{2.0cm}
{\bf J. L. Hewett, H.~Weerts}, 
K. S. Babu, J. Butler, B. Casey, A. de Gouv\^ea,
R.~Essig, Y.~Grossman, D.~Hitlin, J.~Jaros, E.~Kearns, K.~Kumar, Z.~Ligeti,
Z.-T.~Lu, K.~Pitts, 
M.~Ramsey-Musolf, J.~Ritchie, K.~Scholberg, W.~Wester, G.~P.~Zeller

 \tableofcontents

\newpage

\pagenumbering{arabic}


\setcounter{chapter}{1}

\chapter{Intensity Frontier}
\label{chap:IF_start}

\begin{center}\begin{boldmath}

\begin{large}
{\bf Conveners: J. L. Hewett and H.~Weerts}
\end{large}

K.~S. Babu, J.~Butler, B.~Casey, A.~de Gouv\^ea,
R.~Essig, Y.~Grossman, D.~Hitlin, J.~ Jaros, E.~Kearns, K.~Kumar, Z.~Ligeti,
Z.-T.~Lu, K.~Pitts, 
M.~ Ramsey-Musolf, J.~Ritchie, K.~Scholberg, W.~Wester, G.~P.~Zeller

\end{boldmath}\end{center}

\section{Introduction}
\label{sec:IF_intro}

\input IFintro.tex

\section{Neutrinos}
\label{sec:IF_neutrinos}

\input neutrino-summary.tex

\section{Baryon number violation}
\label{sec:IF_bnv}

\input bnv_summary.tex

\section{Charged leptons}
\label{sec:IF_charlept}

\input chargedlepton-summary.tex

\def\K0bar{{\bar{K}}^0}
\def\ProjectX {$Project X$}
\section{Quark flavor physics}
\label{sec:IF_QF}

\input execsumm.tex

\section{Nucleons, nuclei and atoms}
\label{sec:IF_nucnucat}

\input nucleons_summary_short_nofigs.tex

\section{New light, weakly-coupled particles}
\label{sec:IF_NLWCP}

\input NLWCP_summary.tex

\section{Benchmark scenarios}

\input benchmarks.tex

\section{Conclusions}

\input conclusions.tex



\end{document}

%% file: workshopsymbols.tex
\newcommand{\nc}{\newcommand}  

\def\beq{\begin{equation}}
\def\eeq#1{\label{#1}\end{equation}}
\def\eeqn{\end{equation}}

\newenvironment{Eqnarray}%
   {\arraycolsep 0.14em\begin{eqnarray}}{\end{eqnarray}}
\def\beqa{\begin{Eqnarray}}
\def\eeqa#1{\label{#1}\end{Eqnarray}}
\def\eeqan{\end{Eqnarray}}

\nc{\ra}{\rightarrow}  
\nc{\slsh}{\slash\hspace*{-0.22cm}}
\def\Re{{\cal R \mskip-4mu \lower.1ex \hbox{\it e}\,}}
\def\Im{{\cal I \mskip-5mu \lower.1ex \hbox{\it m}\,}}

\nc{\vev}[1]{ \left\langle {#1} \right\rangle }
\nc{\bra}[1]{ \langle {#1} | }
\nc{\ket}[1]{ | {#1} \rangle }
\nc{\fb}{\,{\rm fb}^{-1}}
\nc{\ev}{{\rm eV}}
\nc{\kev}{{\rm keV}}
\nc{\Mev}{{\rm MeV}}
\nc{\gev}{{\rm GeV}}
\nc{\tev}{{\rm TeV}}
\nc{\mev}{{\rm MeV}}

\def\del{\partial}
\def\Dslash{\not{\hbox{\kern-4pt $D$}}}
\def\dslash{\not{\hbox{\kern-2pt $\del$}}}
\def\pslash{\not{\hbox{\kern-2pt $p$}}}
\def\ETmiss{ \not{\hbox{\kern-4pt $E$}}_T }

\def\msb{{\bar{\ssstyle M \kern -1pt S}}}

\def\babar{\mbox{\sl B\hspace{-0.4em} {\small\sl A}\hspace{-0.37em} \sl B\hspace{-0.4em} {\small\sl A\hspace{-0.02em}R}}}

%% file: mainauthorlist.tex
{\large  Study Conveners: M. Bardeen, W. Barletta, L.~A.~T.~Bauerdick, R. Brock,
D.~Cronin-Hennessy, M.~Demarteau, M.~Dine, J.~L. Feng, M. Gilchriese,
S. Gottlieb, J.~L.~Hewett, R. Lipton, H.~Nicholson, M.~E. Peskin,
S. Ritz, I.~Shipsey, H. Weerts}\\
\vspace{1cm}

{\large Division of Particles and Fields Officers in 2013:
J.~L. Rosner (chair), 
I. Shipsey (chair-elect), 
N. Hadley (vice-chair),
P. Ramond (past chair)}\\
\vspace{1cm}

{\large Editorial Committee:
R.~H. Bernstein,
N. Graf,
P. McBride,
M.~E. Peskin,
J.~L. Rosner,
N.~Varelas,
K. Yurkewicz}

%% file: IFintro.tex
Particle physics aims to understand the universe around us.  The Standard Model (SM) of particle physics describes the basic
structure of matter and the forces through which matter interacts, 
to the extent we have been able to probe thus far.  However, it leaves some big questions
unanswered.  Some are within the SM itself, such as why there are so many fundamental particles and why they have
different masses.  In other cases, the SM simply fails to explain some phenomena, such as the observed
matter-antimatter asymmetry in the universe, the existence of dark matter, and the presence of non-zero neutrino masses.
These gaps lead us to conclude that the universe must contain new and unexplored elements of nature.  

These questions are best pursued with a variety of approaches, rather than with a single experiment or technique.
Particle physics uses three basic approaches, often characterized as exploration along the Cosmic, Energy and
Intensity Frontiers.  Each employs different tools and techniques, but they ultimately address the same
fundamental questions.  This allows a multi-pronged approach, where attacking basic questions from different angles
furthers knowledge and provides deeper answers, so that the whole is more than a sum of the parts.

\begin{figure}[ht]
\centerline{\includegraphics[width=1.0\textwidth]{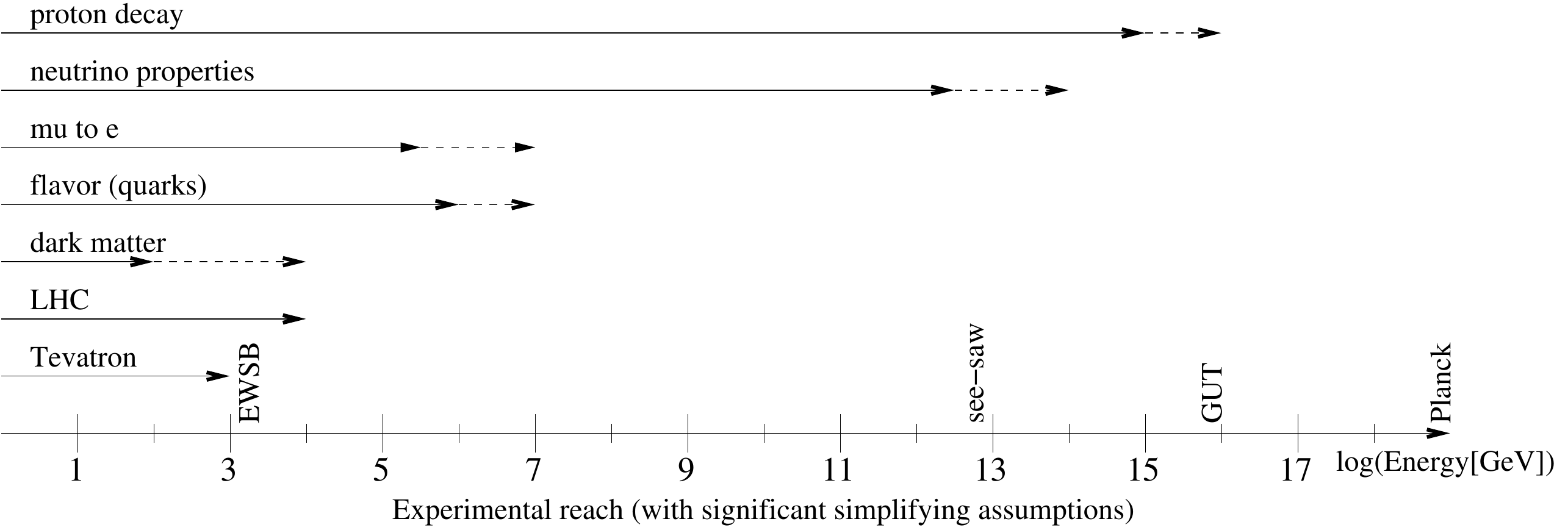}}
\vskip0.1cm
\caption{The energy scale of new physics that is probed by various experimental programs as indicated.  The solid line represents
the present level of experimental sensitivity, while the dashed line indicates the expected sensitivity of proposed facilities.
The Intensity Frontier programs (proton decay, neutrino properties, mu to e, flavor) provide indirect probes of new physics effects.
In contrast, the results presented for the Cosmic (dark matter) and Energy (14 TeV LHC, Tevatron) Frontier programs represent the direct search 
for the production of new particles.
Indirect searches are also possible at these facilities, and increase their sensitivity to high energy scales by roughly an order of 
magnitude. The vertical text shows the energy scale at which Electroweak Symmetry Breaking (EWSB), the neutrino see-saw mechanism, and
Grand Unified Theories (GUT) occur, and where the quantum effects of gravity become strong (Planck).  } 
\label{fig:arrows}
\end{figure}

The Intensity Frontier explores fundamental physics with intense sources and ultra-sensitive detectors.  It encompasses searches 
for extremely rare processes and for tiny deviations from Standard Model expectations.  Intensity Frontier experiments
use precision measurements to probe quantum effects.  They typically investigate new laws of physics that manifest themselves at higher energies
or weaker interactions than those directly accessible at
high-energy particle accelerators. This is illustrated in Fig.~{\ref{fig:arrows}}.
These experiments require the greatest possible beam intensities of neutrinos, electrons, muons, photons or hadrons, as well as 
large detectors.  This provides 
an opportunity for substantial new discoveries complementary to Energy and Cosmic Frontier experiments.

For neutrinos, large underground detectors are central and necessary to understand the physics beyond the Standard Model that 
has been revealed by the flavor oscillation of massive neutrinos.   Coupled with intense neutrino beams, they provide the means to
establish the neutrino mass hierarchy and determine the CP-violating phase inherent in the predominant three-flavor 
paradigm, which is still to be robustly confirmed.  This design also empowers
searches for unexpected phenomena such as non-standard interactions or sterile neutrinos.  Super-massive, underground
detectors continue the search for baryon number violation by increasing the sensitivity to the proton lifetime by an
order of magnitude,
moving deep into the territory motivated by current thinking on the unification of forces.  High-resolution,
low-background underground detectors search for exotic nuclear decays to determine whether neutrinos are
their own antiparticles.

Intense beams of muons are required to search for charged-lepton flavor-violating processes, such as muon to electron conversion, 
$\mu \to e\gamma$ or $\mu \to eee$ decays, which are extremely rare in the Standard Model, but may be detectable in scenarios 
with new physics. Study of the intrinsic properties of muons, such as the magnetic and electric dipole moments, become possible with 
intense muon beams. Large samples of tau leptons, produced at high luminosity $e^+e^-$ storage rings, can search for similar observables 
in the third generation.

In the quark sector, much more precise bottom, charm, and strange particle decay measurements will increase by over an order of 
magnitude the sensitivity to deviations from the Standard Model, as well as allow for the precision measurements of flavor-changing 
neutral-current and CP-violating decays.  This suite of measurements will be crucial to understand the detailed nature of any new 
phenomena the Large Hadron Collider (LHC) at CERN 
may uncover, and will strongly constrain new physics scenarios.

Intense beams of electrons can enable searches for hidden-sector force particles, which may mediate dark matter interactions. Intense 
photon beams may produce para-photons, as well as axions, axion-like particles, or scalars.  Some of these may be connected to dark 
matter or dark energy. The Intensity Frontier can shed light on whole new worlds of light, weakly-coupled particles, which have so 
far remained hidden from observation.

The intensity frontier program was defined in terms of six areas that formed the basis of working groups: experiments
that probe (1) neutrinos\cite{neus}, (2) baryon number violation\cite{bnv}, (3) charged leptons\cite{leptons}, (4) quark flavor\cite{quarks},
(5) nucleons, nuclei, and atoms\cite{nukes}, and (6) new light, weakly-coupled particles\cite{NLWCP}.  The working groups began their efforts
well in advance of the Summer Study, holding regular meetings and soliciting written contributions.  The
2013 Community Summer Study followed
the 2011 workshop on {\it Fundamental Physics at the Intensity Frontier} held in Rockville, MD.  The 2011 workshop was sponsored by the
Office of High Energy Physics in the U.S. Department of Energy Office of Science and was charged with identifying the most compelling
science opportunities at the Intensity Frontier.  The report of the Rockville meeting\cite{Hewett:2012ns} defined the scope of the
Intensity Frontier, which was sustained for the 2013 Summer Study.  An additional resource detailing the piece of the Intensity Frontier program
that is connected to Project X at Fermilab is the report of the 2012 Project X Physics Study\cite{Kronfeld:2013uoa}.

The working group reports\cite{neus,bnv,leptons,quarks,nukes,NLWCP} provide a clear overview of the science program within 
each area of the Intensity Frontier.
The discovery opportunities are presented for facilities that will be available this decade or will come online during
the next decade.  Here, we summarize the findings from each working group from the 2013 Community Summer Study, affectionately called 
``Snowmass''.

%% file: neutrino-summary.tex
\subsection{Overview and summary}
\label{sec:neutrino_summary}

%





Decades of experimental and observational scrutiny have revealed less than a handful of phenomena outside the Standard Model, among them evidence for dark energy and dark matter, and the existence of non-zero neutrino masses.
While many experiments continue to look for other new phenomena and deviations from Standard Model predictions, it is clear that continued detailed study of the neutrino sector is of the utmost importance.    
 
Compared to the other fermions, the elusive nature of the neutrinos has made them extremely difficult to study in detail.    In spite of the challenges, 
\textbf{neutrino physics has advanced quickly and dramatically since the end of the last century}.  Thanks to a remarkable suite of experiments and associated theoretical work, two previously unknown and closely related features of neutrinos now stand out clearly: neutrinos have mass, and neutrinos of different generations mix with each other.
Starting from almost 
no knowledge of the neutrino masses or lepton mixing parameters twenty years ago,
we have built a robust, simple, three-flavor paradigm which successfully describes most of the data. 

Experiments with solar and atmospheric neutrinos, as well as neutrinos produced in reactors and accelerators,
have established, beyond reasonable doubt, that a neutrino produced in a well-defined flavor state (e.g., a muon-type neutrino $\nu_{\mu}$) has a nonzero probability of being detected in a different flavor state  (e.g., an electron-type neutrino $\nu_e$). This flavor-changing probability depends on the neutrino energy and the distance traversed between the source and the detector. The only consistent explanation of nearly all neutrino data collected over the last two decades is a phenomenon referred to as ``neutrino mass-induced flavor oscillation.''

\begin{figure}[ht]
\centerline{\includegraphics[width=1.0\textwidth]{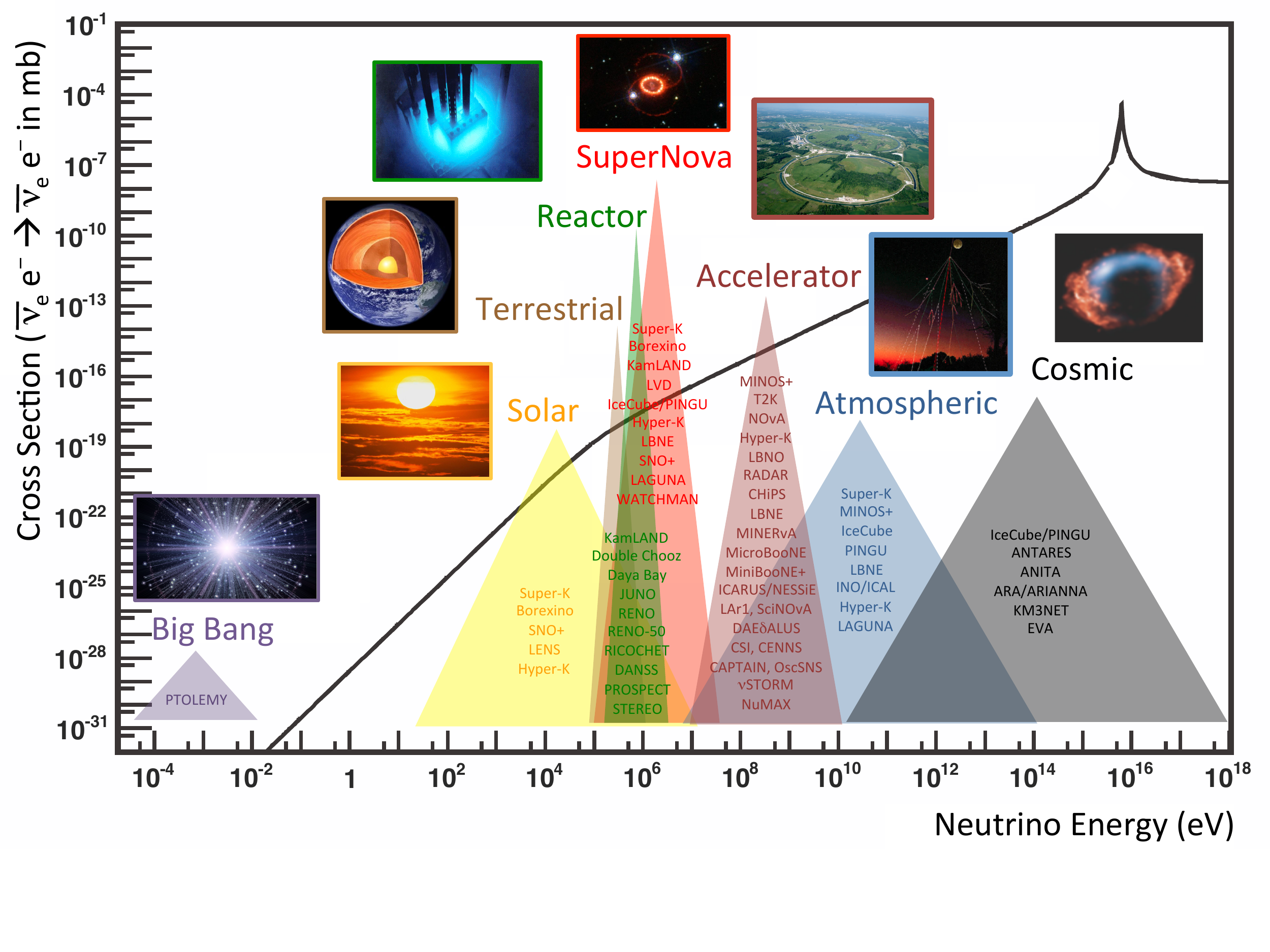}}
\vskip-1.0cm
\caption{Neutrino interaction cross section as a function of energy, showing typical energy regimes accessible by different neutrino sources and experiments. The curve shows the scattering cross section for $\bar{\nu}_e \, e^- \rightarrow e^- \, \bar{\nu}_e$ on free electrons, for illustration. Figure is modified from~\cite{Formaggio:2013kya}}
\label{fig:sources}
\end{figure}

In two different oscillation sectors, similar parallel stories unfolded: Hints of neutrino flavor change in experiments studying  naturally-produced neutrinos were confirmed, and later refined, by experiments with human-made neutrinos.  The disappearance of atmospheric $\nu_\mu$ was unambiguously confirmed by several beam $\nu_\mu$ disappearance experiments, which have now achieved high precision on the driving ``atmospheric'' mixing parameters, \textit{i.e.}, the
mass-squared difference $|\Delta m^2_{32}|$ and the mixing angle $\theta_{23}$. 
The observation of the disappearance of $\nu_e$ from the Sun, a decades-old mystery, was definitively confirmed as evidence for flavor change
using flavor-blind neutral-current interactions.  This ``solar'' oscillation was further confirmed, and the pertinent ``solar'' mixing parameters  ($\theta_{12}$ 
and the mass-squared difference $\Delta m^2_{21}$) were precisely measured using reactor antineutrinos and further solar data.  This complementarity illustrates the importance of exploring the diverse neutrino sources available (see Fig.~\ref{fig:sources}).

The current generation of detectors is now exploring oscillations in a three-flavor context, with both accelerator and reactor tour-de-force experiments having now measured, with good precision, the value of the third mixing angle, $\theta_{13}$, via positive searches for $\nu_\mu \rightarrow \nu_e$ appearance and $\bar{\nu}_e$ disappearance.  


Furthermore, while most of the data fit the three-flavor paradigm very well,
some experiments have uncovered intriguing anomalies  that do not fit this simple picture.
These exceptions include apparent short-baseline  $\nu_\mu\rightarrow\nu_e$ and $\bar{\nu}_\mu\rightarrow\bar{\nu}_e$ transitions, and the anomalous disappearance of reactor and radioactive source electron-type antineutrinos and neutrinos.  Although these hints currently have only modest statistical significance, if confirmed they would be 
evidence for states or interactions present in theories beyond the SM.

The observation of neutrino oscillations implies that neutrinos have nonzero masses, a discovery of fundamental significance.  
We do not know the mechanism responsible for the generation of neutrino masses, but we can state 
with some certainty that new degrees of freedom are required. The number of options is enormous. The current data do not reveal, for example, whether the new physics scale is very low ($\sim 1$~eV) or very high ($\sim 10^{15}$~GeV). The origin of neutrino masses is one of the biggest puzzles in particle physics today, and will only be revealed, and perhaps only indirectly,  with more experimental information from different probes in the different frontiers of particle physics research. Furthermore, the pattern of lepton mixing is very different from that of quarks. We do not yet know what that means, but precision studies of lepton mixing via neutrino oscillations may reveal crucial information regarding the long-standing flavor puzzle.


\subsection{The big questions and physics opportunites}

We are now poised to answer some of the most fundamental and important questions of our time. \textbf{There is a clear experimental path forward}, which builds heavily on the recent successful history of this rapidly-evolving field of particle physics.  We now list these questions in turn and review the proposed experimental program that will address them.

\noindent \textit{What is the pattern of neutrino masses? Is there CP violation in the lepton sector?  To what extent does the three-flavor paradigm describe nature?}

The current neutrino data allow for very large deviations from the three-flavor paradigm. New neutrino--matter interactions as strong as the SM weak interactions are not ruled out.  In fact, mechanisms for generating neutrino mass not only allow,
but sometimes predict, the existence of new ``neutrino'' states with virtually any mass.

Even in the absence of more surprises, we still do not know how the neutrino masses are ordered: 
Do we have two ``light'' and one ``heavy'' neutrino (the so-called normal mass hierarchy) or two ``heavy'' and one ``light'' neutrino (the inverted hierarchy)?
 The resolution of this issue is of the utmost importance, for both practical and fundamental reasons. As will become more clear below, resolving the neutrino mass hierarchy will allow us to optimize the information obtained from other neutrino experimental probes, including searches for leptonic CP-invariance violation, searches for the absolute value of the neutrino masses, and searches for the violation of lepton number via neutrinoless double-beta decay. In addition, the mass hierarchy will also reveal invaluable information concerning the origin of neutrino masses. If the mass hierarchy is inverted, we learn many things. For example, the inverted hierarchy implies that at least two of the three neutrino masses are quasi-degenerate, a condition that is not observed in the spectrum of charged leptons or quarks. 

Experimental neutrino oscillation data revealed that CP invariance can be violated in the lepton 
sector. In particular,
neutrino oscillation studies have the opportunity to explore a brand new source of CP violation. 
The lepton sector accommodates up to three new sources of CP violation in the neutrino mass matrix
-- one Dirac phase, which is analogous to the CKM phase in the quark sector, and two Majorana phases. 
Neutrino oscillations make it possible to explore the Dirac phase.
The Majorana phases are non-zero only if neutrinos are their own antiparticles, and can be deduced 
(depending on the physics of lepton-number violation) from the rate of neutrinoless double-beta decay, a 
determination of the mass hierarchy, and a direct measurement of the neutrino mass. Currently, two sources of 
CP violation are known: the CP-odd phase in the CKM matrix, and the QCD $\bar\theta$ parameter. The former is 
known to be large, while the latter is known to be at most vanishingly small. Exploring CP violation in the lepton 
sector is guaranteed to significantly increase our understanding of this phenomenon. It is also likely that information regarding CP violation in the lepton sector will play a key role when it comes to understanding the mechanism for baryogenesis. 


These questions can only be addressed directly by neutrino oscillation experiments. The current generation of oscillation experiments, including Double Chooz, RENO, Daya Bay, T2K, and NO$\nu$A, will start to resolve the neutrino mass hierarchy. 
Combining their data samples may provide a first glimpse at CP violation in the lepton sector. These will also provide improved measurements of almost all neutrino oscillation parameters. Next-generation experiments, coupled with very intense proton beams, will definitively resolve the neutrino mass hierarchy and substantially improve our ability to test CP invariance in the lepton sector. Next-generation experiments with reactor neutrinos with intermediate baselines (around 50~km), as well as atmospheric neutrino experiments, may also independently resolve the neutrino mass hierarchy. The former will also provide precision measurements of the ``solar'' parameters, $\Delta m^2_{21}$ and $\theta_{12}$. Different experiments with different energies, baselines, and detector technologies will allow for strong tests of the three-flavor paradigm. In the farther future, a more definitive probe of the three-flavor paradigm and precision measurements of CP violation in the lepton sector (or lack thereof) will require long-baseline experiments with different neutrino beams. Leading candidates include 
neutrinos from pion decay at rest produced by high-intensity cyclotron proton sources,
and neutrinos from muon storage rings.
A muon-storage-ring facility should be able to measure the Dirac CP-odd phase with a precision on par with what has been achieved in 
the quark sector, and provide the most stringent constraints on the three-flavor paradigm, thanks to its capability to measure
several different oscillation channels with similar precision.


Comprehensive and detailed studies of neutrino-matter scattering not only serve as tests of the SM and probes of nuclear structure but are also a definite requirement for precision neutrino oscillation experiments.  The convolution of an uncertain neutrino flux with imprecise scattering cross sections and roughly estimated nuclear effects can result in large, even dominant, systematic errors in the measurement of neutrino oscillation parameters.  
More generally, we need to fully characterize neutrino-matter interactions (requiring dedicated experimental and theoretical efforts) to enable deeper understanding of neutrino oscillations, supernova dynamics, and dark matter searches.

\noindent \textit{Are neutrinos Majorana or Dirac particles?}  

Massive neutrinos are special. Among all known fermions, neutrinos are the only ones not charged under the two unbroken gauge symmetries: electromagnetism and color. This implies that, unlike all other known particles, neutrinos may be Majorana fermions. The existence of Majorana neutrinos would imply, for example, that neutrino masses are a consequence of a new fundamental energy scale in physics, potentially completely unrelated to the electroweak scale. Dirac neutrinos, on the other hand, would imply that $U(1)_{B-L}$, or some subgroup, is a fundamental symmetry of nature, with deep consequences for our understanding of the laws of physics.  

If neutrinos are Majorana fermions, lepton number cannot be a conserved quantum number. (Conversely, lepton number violation indicates that massive neutrinos are Majorana fermions.) Hence, the best (perhaps only) probes for the hypothesis that neutrinos are Majorana fermions are searches for lepton number violation. By far, the most sensitive probe of lepton-number conservation is the pursuit of neutrinoless double-beta decay ($0\nu\beta\beta$), $(Z)\to (Z+2)e^-e^-$, where $(Z)$ stands for a nucleus with atomic number $Z$. Independent from the strict connection to the nature of the neutrino, the observation of $0\nu\beta\beta$ would dramatically impact our understanding of nature (similar to the potential observation of baryon number violation) and would provide clues concerning the origin of the baryon asymmetry.  

In many models for the origin of lepton-number violation, $0\nu\beta\beta$ is dominated by the exchange  of virtual massive Majorana neutrinos, in such a way that the amplitude for this process is proportional to 
$m_{ee}\equiv \sum_{i}U_{ei}^2m_i$, $i=1,2,3,\ldots$ under the assumption that all neutrinos are light (which holds for the three SM neutrinos). Under these circumstances, the observation of $0\nu\beta\beta$ would not only reveal that neutrinos are Majorana fermions, it would also provide information concerning the absolute values of the neutrino masses. Conversely, if we know the neutrino mass-squared differences and the magnitude of the relevant elements of the mixing matrix, we can predict the rate for $0\nu\beta\beta$ as a function of the unknown value of the lightest neutrino mass. In particular, if the neutrino mass hierarchy is inverted, there is a lower bound to $|m_{ee}|\gtrsim 20$~meV. 

The current generation of 100-kg-class $0\nu\beta\beta$ search experiments should reach effective masses in the 100~meV range; beyond that, there are opportunities for multi-ton-class experiments that will reach sub-10~meV effective mass sensitivity, pushing below the inverted hierarchy region. In order to fully exploit the relation between $0\nu\beta\beta$ and nonzero Majorana neutrino masses, it is, however, imperative to understand in detail the associated nuclear matrix elements. These require detailed theoretical computations beyond those carried out to date. 

\pagebreak

\noindent \textit{What is the absolute neutrino mass scale?}  

While the values of the neutrino mass-squared differences are known, their absolute mass values remain elusive. In order to properly understand particle physics in general, and neutrinos in particular, it is clear that knowledge of particle masses -- not just mass-squared differences -- is mandatory. The current neutrino data still allow for the possibility that the lightest neutrino mass is vanishingly small, or that the masses of all three known neutrinos  are quasi-degenerate. These two possibilities are qualitatively different and point to potentially different origins for the nonzero neutrino masses.

Neutrino masses can only be directly determined via non-oscillation neutrino experiments. The most model-independent observable sensitive to sub-eV neutrino masses is the shape of the end-point of beta decay spectra. Precision studies of tritium beta decay provide the most stringent bounds, and are expected to play a leading role in next-generation experiments. KATRIN, the most ambitious current-generation  tritium-beta-decay experiment, will directly probe neutrino masses a factor of 10 smaller than the best current bounds. Innovative ideas are needed to go beyond this level of sensitivity.

Other probes of the absolute value of the neutrino masses include $0\nu\beta\beta$, discussed above, and analysis of the large-scale structure of the Universe. Both are, in their own way, much more model-dependent than precision studies of beta decay. Today, cosmological observables provide the most stringent bounds on the absolute values of the neutrino masses, constraining their sums to be below several tenths of an eV. The prospects for the next several years are very exciting.

\noindent \textit{Are there already hints of new physics in existing data?}  

There are intriguing anomalies that cannot be accommodated within the three-flavor paradigm, and suggest new physics beyond it.  In particular, there is marginal ($2-4\, \sigma$) yet persistent evidence for oscillation phenomena at short baselines that is not consistent with the well-established oscillation lengths associated with the ``solar'' and ``atmospheric'' mass-squared differences.  These anomalies, which are not directly ruled out by other experiments, include the excess of $\bar{\nu}_e$ events observed by the LSND experiment, the $\nu_e$ and $\bar{\nu}_e$ excesses observed by MiniBooNE (particularly at low-energies), the deficit of $\bar{\nu}_e$ events observed by reactor neutrino experiments, and the deficit of $\nu_e$ events observed in the SAGE and GALLEX radioactive source experiments. Although there may be several possible ways to explain these anomalies by introducing new physics, the most credible ones, while not ruled out, do not provide a good fit to {\sl all} available  neutrino data. 
Combined, the anomalies are often interpreted as evidence for one or more additional neutrino states, known as sterile neutrinos. 
The $3+N$ sterile neutrino model, in which there are three light mostly-active neutrinos and $N$ mostly-sterile 
neutrinos that mix with the active flavors, is often used to fit the existing data and gauge the reach of proposed next-generation experiments. 
For $N>1$, these models allow for CP-violating effects in short-baseline appearance experiments.  

Beyond particle physics, cosmology hints at the existence of additional neutrinos.  Fits to astrophysical 
data sets -- including the cosmic microwave background, large scale structure, baryon acoustic oscillations, and Big Bang nucleosynthesis -- are sensitive to the effective number of light degrees of freedom ($N_{\rm eff}$).  In the SM,
$N_{\rm eff}$ is equivalent to the effective number of neutrino species, although in principle this could include other types of light, weakly-coupled states.  The recent Planck data are consistent with $N_{\rm eff}=3$ but still allow $N_{\rm eff}=4$. Potential connections between this hint and the short-baseline anomalies above are tantalizing but neither established nor excluded.

All these anomalies in the neutrino sector may disappear with more data; but if they are confirmed, the consequences would open up a whole new sector to explore experimentally and theoretically. The discovery of new neutrino states, for example, would revolutionize our understanding of particle physics. Definitive tests are clearly needed and concrete efforts are already underway. The MicroBooNE experiment, for example, aims at addressing the low-energy excesses observed at MiniBooNE. A variety of neutrino sources and flavor-changing observables  are being pursued as potential means to address the different anomalies. 



\noindent \textit{What new knowledge will neutrinos from astrophysical sources bring?}  

Neutrinos come from natural sources as close as the Earth and Sun, as far away as distant galaxies, and even as remnants from the Big Bang. They range in kinetic energy from less than one meV to greater than one PeV.  As weakly-interacting particles, they probe otherwise inaccessible properties of astrophysical sources; astrophysical neutrino sources shed light on the nature of neutrinos themselves, and on cosmology.  

At the very lowest energies, we can access information about the  $T_{\nu}=1.95$~K Big Bang relic neutrinos via cosmological observables; direct detection of these particles is extremely challenging but nevertheless can be pursued.

In the few to few-tens-of-MeV energy range, 
large underground liquid scintillator, water Cherenkov, and liquid argon detectors are the instruments of choice.
Solar neutrinos may have more to tell us about neutrino oscillations and other neutrino properties, and about solar physics.
Neutrinos from stellar core collapse  have the potential not only to shed light on the astrophysics of gravitational collapse, but also to provide a unique probe of neutrino properties.  It is now even possible to study the Earth via MeV geoneutrinos from terrestrial radioactivity.

In the TeV, and higher, energy region, fluxes of atmospheric neutrinos start to diminish, and neutrinos of astrophysical origin begin to dominate the flux.
Unlike photons and charged particles, these cosmic neutrinos travel unimpeded from their sources and will bring information on the origin of ultra-high-energy cosmic rays.
Due to the very small expected fluxes, the construction of neutrino telescopes requires the instrumentation of enormous natural reservoirs of water.  The recent detection of
the first ultra-high-energy astrophysical neutrinos by IceCube has opened a critical new window of investigation into the study of nature's highest-energy particle
accelerators.

\subsection{The path forward}

Table~2.1 gives a summary of the many current and proposed experiments, in the U.S. and abroad, designed to address the physics questions posed in this section. The number of possibilities is endless, nonetheless we now describe a specific path forward, both in the U.S. and in an international context. 
Neutrino physics is a broad subfield of fundamental particle physics, and requires a multi-pronged approach in order to address all the outstanding questions and fully explore the new physics revealed by neutrino oscillation experiments. 
Investment in a range of large, medium and small-scale neutrino experiments (as well as in detector R\&D and theory) will ensure a healthy program.

\noindent\textit{Comprehensive test of the three-flavor paradigm, via long-baseline, precision neutrino oscillation experiments:} The next-generation experiments will take full advantage of conventional neutrino beams from pion decay in flight. These will begin to over-constrain the parameter space, and will start to seriously explore CP-violating phenomena in the lepton sector.
The U.S., with the Long-Baseline Neutrino Experiment (LBNE) and a future multi-megawatt beam from Project
X, is uniquely positioned to lead an international campaign to measure
CP violation and aggressively test the three-flavor paradigm.
Complementary experiments with different energies, baselines, and detector technologies ({\it e.g.}, Hyper-K in Japan) are required in order to fully exploit conventional neutrino beams. 
The accompanying very-large detectors, if placed underground, also allow for the study of atmospheric neutrinos, nucleon decay, and precision measurements of neutrinos from a galactic supernova explosion.
PINGU, an upgrade of IceCube, provides a promising opportunity to measure the mass hierarchy using atmospheric neutrinos. 

Further in the future, experiments will require more intense, and better understood neutrino beams.  Promising possibilities include neutrinos from muon storage rings (e.g., NuMAX), and neutrinos from very intense cyclotron-based sources of pion decay at rest (e.g., DAE$\delta$ALUS). 
Muon-based neutrino beams in particular have strong synergies with
Project X and provide a necessary step in the R\&D for a high-energy muon collider. 
 While these large, ambitious projects are vigorously developed, the following medium and small-scale neutrino activities need to be pursued.
\begin{itemize}
\item \textit{Precision measurement of neutrino cross sections and a detailed understanding of the neutrino flux from pion-decay-in-flight neutrino beams.} These activities can be pursued in the ``near-detectors'' associated with the large, long-baseline projects or alongside R\&D projects related to next-next generation neutrino beams, as well as by small-scale dedicated experiments. A well-considered program of precision scattering experiments in both low- and high-energy regimes, combined with a renewed dedicated theoretical effort to develop a
reliable, nuclear-physics based description of neutrino interactions in nuclei is mandatory.  Scattering measurements may also be of intrinsic interest. 
\item \textit{Definite resolution of the current short-baseline anomalies.} These will (probably) require neutrino sources other than pion decay in flight and the pursuit of different flavor-changing channels, including $\nu_{e,\mu}$ disappearance and $\nu_{\mu}\to\nu_e$ appearance, using a combination of reactor, radioactive source, and accelerator experiments.  In addition to small-scale dedicated experiments, such experiments can be carried out as part of  R\&D projects related to next-next generation neutrino beams (e.g., nuSTORM, IsoDAR).
\item \textit{Vigorous pursuit of R\&D projects related to the development of next-next generation neutrino-beam experiments.} Medium and small experiments associated with such projects could also address several key issues in neutrino physics.
\end{itemize} 

\noindent\textit{Searches for neutrinoless double-beta decay:} The current generation of experiments is pursuing various detector technologies with different double-beta decaying isotopes. The goals of these experiments are to (1) discover neutrinoless double-beta decay, which is guaranteed if the neutrinos are Majorana fermions and their masses are quasi-degenerate 
({\it i.e.}, masses of order 0.1-0.5 eV), and (2) provide information regarding the most promising technique for the next-generation of ton-scale experiments. Next-generation experiments aim at discovering neutrinoless double-beta decay if neutrinos are Majorana fermions and if the neutrino mass hierarchy is inverted. In the case of a negative result, assuming oscillation experiments have revealed that the neutrino mass-hierarchy is inverted, these experiments will provide strong evidence that the neutrinos are Dirac fermions. As with precision measurements of beta decay (see below), the information one can extract from the current and the next generation of neutrinoless double-beta decay experiments increases significantly if indirect evidence for neutrino masses is uncovered, for example, with cosmological probes. 

\noindent\textit{Determination of the absolute values of the neutrino masses:} Precision measurements of beta decay remain the most promising model-independent probes. While the KATRIN experiment is taking data, vigorous R\&D efforts for next-generation probes (e.g., ECHo, Project 8, PTOLEMY) are required in order to identify whether it is possible to reach sensitivities to the effective ``electron-neutrino mass'' below $0.05$~eV. Nontrivial information is expected from different cosmological probes of the large-scale structure of the Universe. 

The allure and relevance of neutrino science and technology extends well beyond the fundamental research community.  The neutrino signal itself may be useful for monitoring reactors in the context of international nuclear nonproliferation.  The essential building blocks of neutrino science  --  detectors  and accelerators  --  have important spin-off applications for medicine and in industry. Finally, ever since neutrinos were first postulated and discovered, their unusual, ghostlike properties and non-intuitive behavior have fascinated the general public. The success of our field depends in no small part on our ability to effectively  convey both the mystery and utility of neutrino science to the public,  Congress, policy-makers, and  funding agencies.

The diversity of physics topics that can be probed through the neutrino sector is enormous and the interplay between neutrino physics and other fields is rich.   Neutrinos have and will continue to provide important information on structure formation in the early universe, Earth, solar and supernova physics, nuclear properties, and rare decays of charged leptons and hadrons. Conversely, information regarding neutrino properties and the origin of neutrino masses  is expected from the Energy and Cosmic Frontiers, and from other areas of Intensity Frontier research. 
In other words, the neutrino sector sits at the nexus of the worldwide effort in Energy, Intensity and Cosmic Frontier physics.


\footnotesize
\begin{table}[p]
\footnotesize
\label{table:experimentstable}
\caption{Summary of the very many current and proposed experiments, in the U.S. and abroad, designed to address various physics questions. Rows refer to neutrino sources and columns
refer to categories of physics topics these sources can address.
The intent is not to give a ``laundry list'', but to give a sense of the activity and breadth of the field.  Some multipurpose experiments appear under more than one physics category.
Experiments based in the U.S. (or initiated and primarily led by U.S. collaborators) are shown in blue and underlined  (note that many others have substantial U.S. participation or leadership).  Proposed and future experiments are in bold; current experiments (running or with construction well underway) are in regular font. More details and references can be found in the subsections of the Neutrino Working Group report \cite{neus}.  Below, DAR refers to decay at rest and DIF to decay in flight.}
\begin{center}
\begin{tabular}{|c|c|c|c|c|c|c|c|} \hline
Source & 3-flavor osc.& Maj./Dirac & Abs. Mass & Interactions & Anomalies/Exotic$^2$ & Astro/Cosmo \\  \hline \hline
Reactor  & KamLAND,& && {\color{blue}\textbf{\uline{RICOCHET}}}& \textbf{DANNS}, \textbf{STEREO}  & \\ 
  & Double Chooz,&&& & {\color{blue}\textbf{\uline{PROSPECT}}}   &  \\
           & Daya Bay, \textbf{JUNO}, & & & &  {\color{blue}\textbf{\uline{RICOCHET}}}  & \\ 
            & RENO, \textbf{RENO-50}&&&& {\color{blue}\textbf{\uline{US Reactor}}} & \\ \hline   
Solar & Super-K, &&&&& Super-K,\\ 
              & Borexino, SNO+, &&&& & Borexino, SNO+,\\
            & \textbf{Hyper-K}, {\color{blue}\uline{\textbf{LENS}}} &&&&& \textbf{Hyper-K}, {\color{blue}\uline{\textbf{LENS}}}\\  \hline
Supernova$^1$  & Super-K, Borexino, &&&&  & Super-K, Borexino, \\  
            & KamLAND, LVD, &&&& & KamLAND, LVD\\
            & {\color{blue}\uline{IceCube/\textbf{PINGU}}},&&&&&{\color{blue}\uline{IceCube/\textbf{PINGU}}},\\
             & \textbf{Hyper-K}, {\color{blue}\uline{\textbf{LBNE}}},  &&&&& \textbf{Hyper-K}, {\color{blue}\uline{\textbf{LBNE}}},\\
            & SNO+, \textbf{LAGUNA},  &&&&& SNO+, \textbf{LAGUNA},\\
             & {\color{blue}\uline{\textbf{WATCHMAN}}} &&&&& {\color{blue}\uline{\textbf{WATCHMAN}}}\\ \hline
Atmospheric&Super-K, {\color{blue}\uline{MINOS+}} , &&&& &\\ 
& {\color{blue}\uline{IceCube/\textbf{PINGU}}}, &&&& &\\  
& {\color{blue}\uline{\textbf{LBNE}}}, \textbf{ICAL}, &&&&&\\ 
&\textbf{Hyper-K},  &&&&&\\ 
&\textbf{LAGUNA} &&&&&\\ \hline 
Pion DAR & {\color{blue}\uline{\textbf{DAE$\delta$ALUS}}} &&&{\color{blue}\uline{\textbf{OscSNS}}},{\color{blue}\uline{\textbf{CSI}}}& {\color{blue}\uline{\textbf{OscSNS}}}&\\ 
 &  &&& {\color{blue}\uline{\textbf{CENNS}}}, & &\\ 
 &  &&& {\color{blue}\uline{\textbf{CAPTAIN}}} & &\\ \hline  
Pion DIF  &  {\color{blue}\uline{MINOS+}}, T2K, &&&{\color{blue}\uline{MicroBooNE}},&{\color{blue}MicroBooNE},&\\  
  &  {\color{blue}\uline{NO$\nu$A}}, \textbf{Hyper-K},&&&{\color{blue}\uline{{MINER$\nu$A}}} & {\color{blue}\uline{\textbf{MiniBooNE+/II}}},&\\  
  & \textbf{LAGUNA-LBNO}, &&&{\color{blue}\uline{No$\nu$A}}&\textbf{Icarus/NESSiE},&\\   
   &  {\color{blue}\uline{\textbf{RADAR}}}, {\color{blue}\uline{\textbf{CHIPS}}}, &&&{\color{blue}\uline{SciNO$\nu$A}} & {\color{blue}\uline{\textbf{LAr1}}},{\color{blue}\uline{\textbf{LAr1-ND}}} &\\  
   &  {\color{blue}\uline{\textbf{LBNE}}},\textbf{ESS$\nu$SB} &&&&{\color{blue}\uline{{MINOS+}}}&\\  \hline
$\mu$DIF & {\color{blue}\uline{\textbf{NuMAX}}} & & & {\color{blue}\uline{\textbf{nuSTORM}}} &{\color{blue}\uline{\textbf{nuSTORM}}} &  \\ \hline
Radioactive & & Many: see& KATRIN,&& \textbf{SOX}, \textbf{CeLAND},&\\ 
Isotopes &  & Nu2 report  &  {\color{blue} \textbf{\uline{Project 8}}},&& \textbf{Daya Bay Source}, &\\  
 &  & for table &  \textbf{ECHo}, &&  \textbf{BEST}  &\\  
 &  &  & {\color{blue}\uline{\textbf{PTOLEMY}}} & &{\color{blue}\uline{\textbf{IsoDAR}}}&\\  \hline  
Cosmic neutrinos & & & & & & {\color{blue}\uline{IceCube/\textbf{PINGU}}}, \\ 
 & & & & & &  ANTARES/\textbf{ORCA}, \\ 
 & & & & & &  {\color{blue}\textbf{\uline{ARA},\uline{ARIANNA}}}, \\
 & & & & & &  {\color{blue}\uline{ANITA}, \uline{\textbf{EVA}}}, \\
& & & & & &  \textbf{KM3NET} \\ \hline
\end{tabular}
\end{center}
\footnotesize
\vspace{-0.11in}
$^1$Included are only kton-class underground detectors; many others would also record events.  
$^2$We note that nearly all experiments can address anomalies at some level; we include in this column only those with this as a primary physics goal.
\end{table}
\normalsize

%% file: bnv_summary.tex

\subsection{Overview and summary}
\label{sec:sum-bnv}




The stability of ordinary matter is attributed to the conservation of
baryon number, $B$. The proton and the neutron are assigned $B =
+1$, while their antiparticles have $B=-1$, and the leptons and
antileptons all have $B=0$. However, many leading theoretical
ideas suggest that baryon number may be only an approximate symmetry,
one that is violated by small amounts. Experimental observation of
baryon number violation would have a profound impact on our
understanding of the evolution of the Universe and the basic forces of
nature. Baryon number violation is an essential ingredient for the
creation of the observed asymmetry of matter over antimatter from the
symmetrical Universe that emerged from the Big Bang. And baryon number
violation is a hallmark prediction of grand unified theories.

\subsection{Theoretical motivation}  

Grand unified theories (GUTs) are well-motivated
on several grounds. They unify the strong, weak and
electromagnetic forces into a single unified force.  There is
evidence from data on the force couplings at low energies that
these couplings may unify to a single value at high energies.
The three
gauge couplings of the SM appear to take on a common value 
at a scale of about $10^{16}$ GeV,
depending on the new physics that lies at the intermediate energies
above the electroweak scale. Such a large energy scale (amounting to
distance scale of order $10^{-30}$ cm) is well beyond the reach of any
conceivable particle accelerators.

GUTs are based on symmetries and organize quarks and leptons into simple unified
representations of symmetry groups that naturally explain the equality of the magnitude
of the electric charge of the electron to that of the proton. GUTs
provide an understanding of the quantum numbers of the quarks and
leptons, as well as an explanation for the miraculous cancellation of
chiral anomalies within each generation. GUTs,
notably those based on the gauge symmetry SO(10), require
the existence of right-handed neutrinos, one per family, which enable
the seesaw mechanism to generate small Majorana masses for the
ordinary neutrinos.  These right-handed neutrinos are prime candidates
for the generation of the baryon asymmetry of the Universe via
leptogenesis.

\begin{figure}[h!]
\begin{center}
\includegraphics[width=0.85\textwidth]{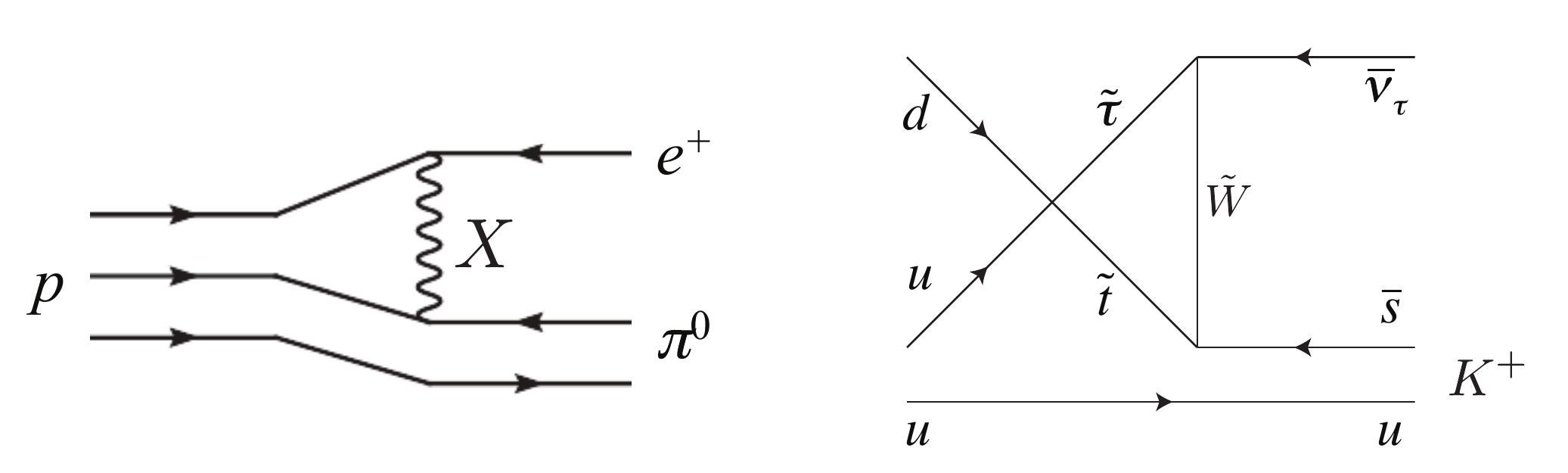}
\caption{\label{pdk-diag} Diagrams inducing proton decay in SUSY GUTs.
  $p \rightarrow e^+ \pi^0$ mediated by $X$ gauge boson (left), and $p
  \rightarrow \overline{\nu} K^+$ generated by a $d=5$ operator.
  (right).}\label{fig:unif-diag}
\end{center}
\end{figure}

Grouping of quarks with leptons, and particles with antiparticles,
into a common GUT multiplet leads to the violation of baryon number
and thus proton decay. Super-heavy so-called $X$ and $Y$ gauge bosons mediate
proton decay through dimension-6 operators, as illustrated in the left
panel of Fig.~\ref{fig:unif-diag}. The minimal version of the SU(5)
GUT, the first GUT model, predicted proton decay within reach of the first-generation proton
decay experiments.  This model has already been excluded by
their experimental limits on $p \rightarrow e^+ \pi^0$, as well as the
mismatch of the three gauge coupling unification when extrapolated to scales of
$10^{14}$ to $10^{15}$ GeV. GUTs based on larger symmetries such as SO(10) are consistent with
both gauge coupling unification and experimental constraints, particularly if they include
supersymmetry, as discussed below.  Other possible
theories include flipped SU(5), which favors the second generation
and predicts the decay $p \rightarrow \mu^+ \pi^0$, and higher-
dimensional GUTs, including those where quarks and leptons live on
separate branes. Even negative experimental results in the search for
proton decay provide valuable information to the pursuit of realistic grand
unified theories.

Supersymmetric grand unified theories (SUSY GUTs) are natural
extensions of the SM that preserve the attractive features
of GUTs, and predict a more precise unification of the three
gauge couplings. The unification occurs at the higher energy of $2
\times 10^{16}$ GeV, suppressing the dimension-6 gauge mediation
responsible for $p \rightarrow e^+ \pi^0$ to a lifetime of at least a
few $\times 10^{34}$ years, compatible with current experimental limits.  
Lifetimes of this magnitude are
now being probed by Super-Kamiokande.  However, the predictions are
uncertain by an order of magnitude or so, due to the unknown masses of
supersymmetric particles and GUT scale particles.

Supersymmetric GUTs (SUSY GUTs) bring a new twist to proton decay, by predicting the decay
mode $p \rightarrow \overline{\nu} K^+$ which is mediated by a colored Higgsino
generating a dimension-5 operator, as shown in the right panel of
Fig.~\ref{fig:unif-diag}. Here, the $d=5$ operator involves the
electroweak wino and supersymmetric partners of the quarks and leptons. The
predicted lifetime for this mode in minimal supersymmetric
SU(5) theories is typically less than $10^{32}$ years, much shorter than the
current experimental lower limit of $6 \times 10^{33}$ years, provided
that the supersymmetric particle masses are less than about 3 TeV. 
This limit is difficult to avoid in
minimal SUSY SU(5) theories unless the supersymmetric particle masses are
much above 3 TeV. 
However, there are 
non-minimal supersymmetric SU(5) models, as well as SUSY GUTs based on the larger symmetry
SO(10), that accommodate the current experimental bounds and predict
proton decay within reach of current and next generation experiments.
SUSY GUTs generally prefer decays into strange mesons, and although $p
\rightarrow \overline{\nu} K^+$ is predominant, modes such as $p
\rightarrow \mu^+ K^0$ are sometimes favored.

One class of minimal SO(10) models which employs a single representation that
contains the Higgs boson 
has been developed. Owing to their minimality,
these models are quite predictive with regard to the neutrino mass
spectrum and oscillation angles. Small quark mixing angles and large
neutrino oscillation angles emerge simultaneously in these models at the weak scale,
despite their parity at the fundamental level.  The neutrino
oscillation angle $\theta_{13}$ is predicted to be large in these
models. In fact, this mixing angle was predicted to have the value
$\sin^22\theta_{13} \simeq 0.09$, well before it was measured to have
this central value. Proton decay studies of these models in the supersymmetric
context show that at least some of the modes among $p \rightarrow
\overline{\nu} \pi^+$, $n \rightarrow \overline{\nu} \pi^0$, $p
\rightarrow \mu^+ \pi^0$ and $p \rightarrow \mu^+ K^0$ have inverse
decay rates of order $10^{34}$ years, while for $p \rightarrow e^+
\pi^0$ the rate is of order $10^{35}$ years.

A common prediction of many GUTs is that proton decay modes obey
the selection rule $\Delta (B-L) =0$, where $L$ denotes lepton number.
For example, the proton decay channels $p \rightarrow e^+ \pi^0$
and $p \rightarrow \overline{\nu} K^+$ follow this rule. 
However, theories with additional spatial dimensions  
predict that $|\Delta(B-L)| =2$, which leads to decays such as $n
\rightarrow e^-+\pi^+$ and $n \rightarrow e^- K^+$. 
Other possible
operators may result in three-body proton decay modes such as $p \rightarrow
l^+\overline{\nu}\nu$ or $p \rightarrow l^+\pi^0\pi^0$. A thorough
search for nucleon decay should consider a multitude of possible final
states. However, $p \rightarrow e^+\pi^0$ and $p \rightarrow
\overline{\nu} K^+$ serve as adequate benchmarks to evaluate future
experiments, since they are routinely predicted to be the dominant decay modes, and the
detector features that efficiently find these modes will usually
perform well when searching over a wide array of modes.

\subsection{Experimental programs}

The Super-Kamiokande (Super-K) water Cherenkov experiment dominates the
current limits set on the lifetime of the proton and bound neutron. No
signs of nucleon decay have been seen yet in any Super-K analysis of
any nucleon decay mode. The 22,500-ton fiducial mass of the detector
contains $7.5\times10^{33}$~protons and $6.0\times10^{33}$~neutrons. Fully
contained atmospheric neutrino interactions in the GeV range
constitute the background. The background is estimated 
using the detailed model
that is standard for Super-K, one that takes into account the
atmospheric neutrino flux, neutrino-nucleus cross sections, and
interactions within the nucleus. The proton decay signal is simulated
using a Monte Carlo program that includes effects due to Fermi motion,
nuclear binding potential, intranuclear reactions, and correlated
nucleon effects (all of which are absent for the free proton that is
the hydrogen nucleus). For the two benchmark modes mentioned above,
Super-Kamiokande has observed no candidates for proton decay and reported the following
90\% C.L. on the partial lifetimes:
\begin{equation}
 \tau(p \rightarrow e^+\pi^0) > 1.4 \times 10^{34} \; {\rm yrs}, 
 \;\;\;\;\;\;\;\;\;
 \tau(p \rightarrow \bar \nu K^+  )>  5.9 \times 10^{33} \; {\rm yrs}.
\end{equation}
\noindent The Super-K collaboration expects to operate the detector
for many more years, as the far detector for T2K and as a supernova
detector, but also to continue the search for proton decay.

A next-generation underground water Cherenkov detector, 
Hyper-Kamiokande (Hyper-K), is proposed in the Kamioka
mine in Japan. It will serve as the far detector of a long baseline
neutrino oscillation experiment using the off-axis J-PARC neutrino
beam.  It will also be capable of observing nucleon decays,
atmospheric neutrinos, and neutrinos of astronomical origin. The
fiducial mass of the detector is planned for 0.56 million metric tons,
and 99,000 photomultipliers would instrument the detector with photon
coverage of 20\% (but probably with enhanced quantum efficiency
compared to the PMTs used in Super-K). 

The sensitivity of Hyper-K for nucleon decays has been studied based
on scaling of the Super-K analysis. Due to the very large
mass, Cherenkov ring reconstruction in the GeV range,
and demonstrated low background rate, water Cherenkov is the best
detector technology for the proton decay signatures $e^+\pi^0$ and
$\mu^+\pi^0$. It is safe to assume an overall signal efficiency of
40\%, where the inefficiency is dominated by the nuclear interaction of
the pion. For comparison, the detection efficiency for the decay of a free
proton in water is 87\%. The background rate is well-established by
Monte Carlo studies, plus measurements using the T2K near detector; one
can conservatively assume two events per megaton-years. Based on these
numbers, the 90\% C.L. sensitivity of Hyper-K for a 10-year exposure
is greater than $10^{35}$ years.

For $p \rightarrow \nu K^+$, most sensitivity comes from the two
relatively background-free decay modes: $K^+ \rightarrow \pi^+\pi^0$
and $K \rightarrow \mu^+\nu$ with nuclear-$\gamma$ tag. Based on the estimated
background rate for Super-K, a 10-year exposure of
Hyper-K would have an expected background between 20 and 35
events with an efficiency in the range 13\% to 19\%. If the detected
number of events is equal to the background expectation, the 90\%
CL limit would be roughly $3 \times 10^{34}$ years for this mode. This estimation
assumes that a reoptimization of the analysis ({\it i.e.,} tighter cuts) to accommodate
the higher background rate will not be performed.

The ability to fully reconstruct particle interactions in high resolution
has long been recognized as the selling point for the liquid argon
time projection chamber (LArTPC) technology. Resolving neutrino
interactions was the prime motivation for selecting the LArTPC technology for
the far detector of the LBNE long-baseline neutrino oscillation
experiment. Using this detector to search for proton decay (and
other non-accelerator physics) is the primary motivation for locating LBNE deep
underground. The fine spatial resolution of the LArTPC coupled with a $dE/dx$
measurement allows the reconstruction of all final-state charged
particles, including an accurate assessment of particle type, by
distinguishing between muons, pions, kaons, and
protons. Electromagnetic showers are also readily measured with a
significant ability to distinguish between those that originate 
from $\pi^0$ decay into photons and those that originate from charged-current
electron neutrino interactions. 

Per kiloton, LArTPC
technology will outperform water Cherenkov in both detection
efficiency and atmospheric neutrino background rejection for most
nucleon decay modes, although intranuclear effects are smaller for
oxygen and non-existent for hydrogen. Taking mass and cost into
account, however, water Cherenkov technology still performs the best for the $p\to
e^+\pi^0$ final state topology, where the signal efficiency is roughly
40\% and the background rate is two events per megaton-year. The corresponding
estimate for a LArTPC is 45\% efficiency and one event per megaton-year,
which is not enough of an improvement to overcome the penalty of lower
mass.

For the $p \rightarrow \nu K^+$ channel, the final-state kaon
is below Cherenkov threshold, the signal efficiency is less than 20\%,
and the background rate is substantial. This mode is thus well-suited to a
LArTPC, where the $K^+$ track is reconstructed with the correct
specific ionization and range. The decay products of the kaon are also
identified with high efficiency for all kaon decay channels. The
total efficiency for the $\nu K^+$ mode is estimated to be as high as
96.8\%, with a background rate of one event per megaton-year. Based on
these numbers and a ten-year exposure, the 34 kiloton LBNE and
560 kiloton Hyper-K detectors have comparable sensitivity (at 90\% C.L.),
but the LArTPC would have an estimated background of 0.3 events
compared to tens of events for Hyper-K. Experimental searches for rare
events in the presence of significant backgrounds are notoriously more
problematic than background-free searches. A 10-year exposure of the
34-kiloton LBNE LArTPC detector would set a partial lifetime limit of
roughly $3 \times 10^{34}$ years, nearly the same as the sensitivity
of Hyper-K, despite 20 times fewer proton-years. There are a number of
other nucleon decay modes for which a LArTPC competes well with water
Cherenkov, mainly because of high background rates in the water
Cherenkov analysis.

\begin{figure}[h!]
\begin{center}
\includegraphics[width=0.90\textwidth]{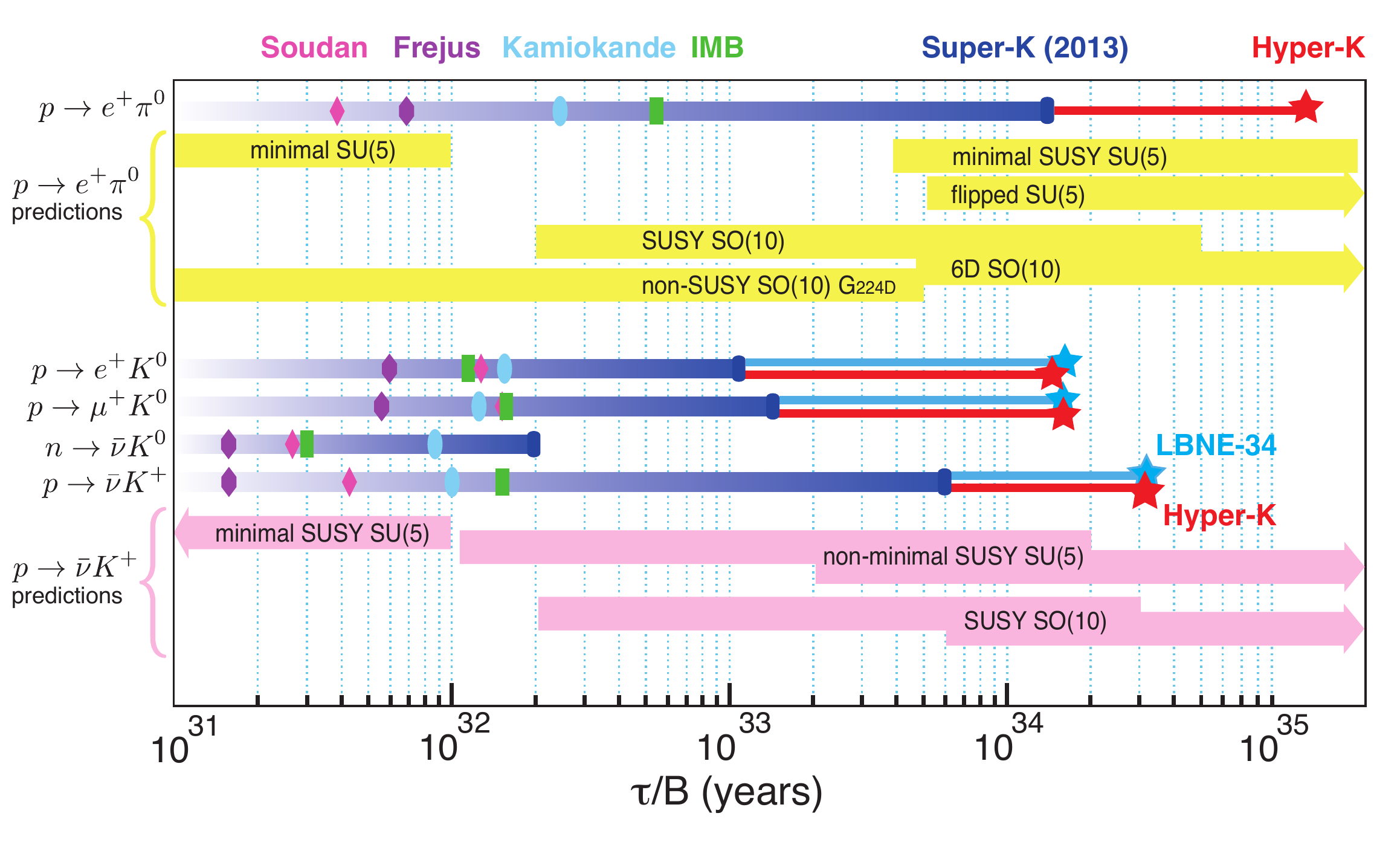}
\caption{\label{fig:pdk-reach} The experimental reach of
  Hyper-Kamiokande and a 34-kton LBNE LArTPC are compared to prior
  experiments and the rough lifetime predictions from a wide range of
  GUT models. The projected limits are for 10 live years of running,
  at 90\% C.L., calculated for a Poisson process including background
  assuming the detected events equal the expected background.}
\end{center}
\end{figure}

Figure~\ref{fig:pdk-reach} shows past experimental limits on 
various proton decay modes from first
generation experiments, the current lifetime limits from Super-K, and
projected lifetime limits for ten years of running of Hyper-K
and a 34-kiloton LBNE LArTPC. These are compared with theoretical predictions from
the various GUTs mentioned
above, where the depicted bands roughly capture the expected ranges.
The next generation experiments Hyper-K and LBNE will make
substantial progress beyond the Super-K search and are sensitive to
lifetimes predicted by numerous well-motivated GUT models.

\subsection{Neutron-antineutron oscillation}

Grand unified theories and the search for nucleon decay, as described
above, are the principal tools for understanding baryon number
violation by one unit. However, baryon number may be violated by two
units through processes such as dinucleon decay and neutron-antineutron
oscillation and is predicted at observable rates in other classes of models. 
These processes afford violation of baryon
number without violation of lepton number. They arise from 
operators with new physics contributions at mass scales of $100-1000$ TeV, well
below the GUT scale. Interesting theoretical models, especially those based
on quark-lepton symmetry realized at the multi-TeV scale, predict a
$n-\overline{n}$ oscillation time of order $10^{10}$ sec. 
Observation of $n-\bar{n}$ oscillations with probability
not far from the current limit could refute the paradigm of
leptogenesis, and in itself would suggest an alternate source
for generating the baryon asymmetry of the universe.

One search technique for neutron-antineutron oscillation 
involves a beam of free neutrons where one neutron could
transform into an antineutron and annihilate in a distant detector.
The probability of oscillation is given by $P(n \rightarrow \bar{n},t)
\simeq [\delta m \, t]^2$, where $\delta m$ is the baryon-violating
interaction strength. The same process may also occur for 
nuclei, where bound neutrons may transform into antineutrons and
annihilate within the nucleus, producing an isotropic final state of
pions. In the case of bound neutrons, a nuclear suppression factor
must be applied to calculate a free oscillation
probability. Currently, the limits from bound $n - \overline{n}$
transformation in oxygen nuclei in Super-Kamiokande are slightly more
restrictive than the result from the best free neutron oscillation
search.  However, the sensitivity in water is hindered by a large background
rate of roughly 25 events per 100 kiloton years with 12\% signal
efficiency. It is hoped that a LArTPC will provide better
performance, but this is still under study.

The previous experimental search for free $n-\bar{n}$ transformations
employed a cold neutron beam from the research reactor at Institute
Laue-Langevin (ILL) in Grenoble, and reported a limit on the transformation
of $\tau>8.6 \times
10^7$ seconds. The average velocity of the cold neutrons was roughly
$600$ ${\rm m/s}$ and the neutron observation time was approximately 0.1
seconds. A net magnetic field of less than 10 nT was maintained over
the evacuated flight volume in order to satisfy the quasi-free
conditions required for oscillations to occur.  Antineutron appearance
was sought through annihilation in a $130 \mu$m carbon film
target, generating a star pattern of several secondary pions, viewed
by a tracking detector, and an energy deposition of 1-2 GeV in the
surrounding calorimeter. In one year of operation this ILL experiment
saw zero candidate events.

A new experimental search for oscillations using free neutrons from a
1 MW spallation target, NNbarX at Project X, has the potential to improve
existing limits on the free oscillation probability by at least 2 orders of
magnitude (and potentially up to 4 orders of magnitude with further R\&D)
by exploiting a new slow neutron source together with an optics
technology developed for materials science experiments. The possibility to create
a dedicated neutron source for particle physics experiments offers a
unique opportunity to design a neutron optical system that accepts a
large fraction of the neutron flux. The initial goal of NNbarX will be
to improve on the probability of an $n-\bar{n}$ transition
by at least a factor 30 compared to the previous
limit set in the ILL-based experiment. This is achievable with existing
technology. A substantial feature of the experiment is the opportunity
to validate a discovery by removing the required magnetic field
shielding, destroying the quasi-free condition, and eliminating the
signal events. The R$\&$D phase of the experiment, including
development of the conceptual design of the cold neutron spallation
target, and conceptual design and optimization of the performance of
the first-stage of NNbarX, is expected to take two to three years. Preliminary
results from this effort suggest that an improvement over the ILL
experiment by a factor of more than 100 may be realized in the
horizontal mode. A vertical flight path currently
under study could achieve further increases in sensitivity. 
The running time of the first stage of
NNbarX experiment is anticipated to be 3 years. Future stages of
NNbarX will depend upon the demonstration of the
technological principles and techniques of the first stage.

%% file: chargedlepton-summary.tex

The  charged
lepton experimental program has enormous physics potential. There are discovery opportunities in experiments that will be conducted over the coming decade using existing facilities, and in more sensitive experiments possible with future facilities such as Project X.
Exquisitely sensitive searches for rare decays of muons and tau leptons, together with precision measurements of their properties, will either elucidate the scale and dynamics of flavor generation, or
limit the scale of flavor generation to well above $10^4$ TeV.  

\subsection{Flavor-violating processes}

The crown jewel of the program is the discovery potential of muon and tau decay experiments searching for charged lepton flavor violation (CLFV) with several orders-of-magnitude improvement in sensitivity in
multiple processes.  An
international program of CLFV searches exists, with experiments recently completed, currently running, and
soon to be constructed in the United States, Japan, and Europe.  These include the completion of the MEG experiment at PSI, an upgrade of MEG,  the proposed mu3e search at PSI, new searches of muon to electron conversion (Mu2e at Fermilab, COMET at J-PARC), studies of $\tau$ decay at SuperKEKB, and over the longer term, experiments exploiting megawatt proton sources such as Project X.

\begin{figure}[ht]
\begin{minipage}[b]{0.48\linewidth}
\centering
\includegraphics[width=\linewidth]{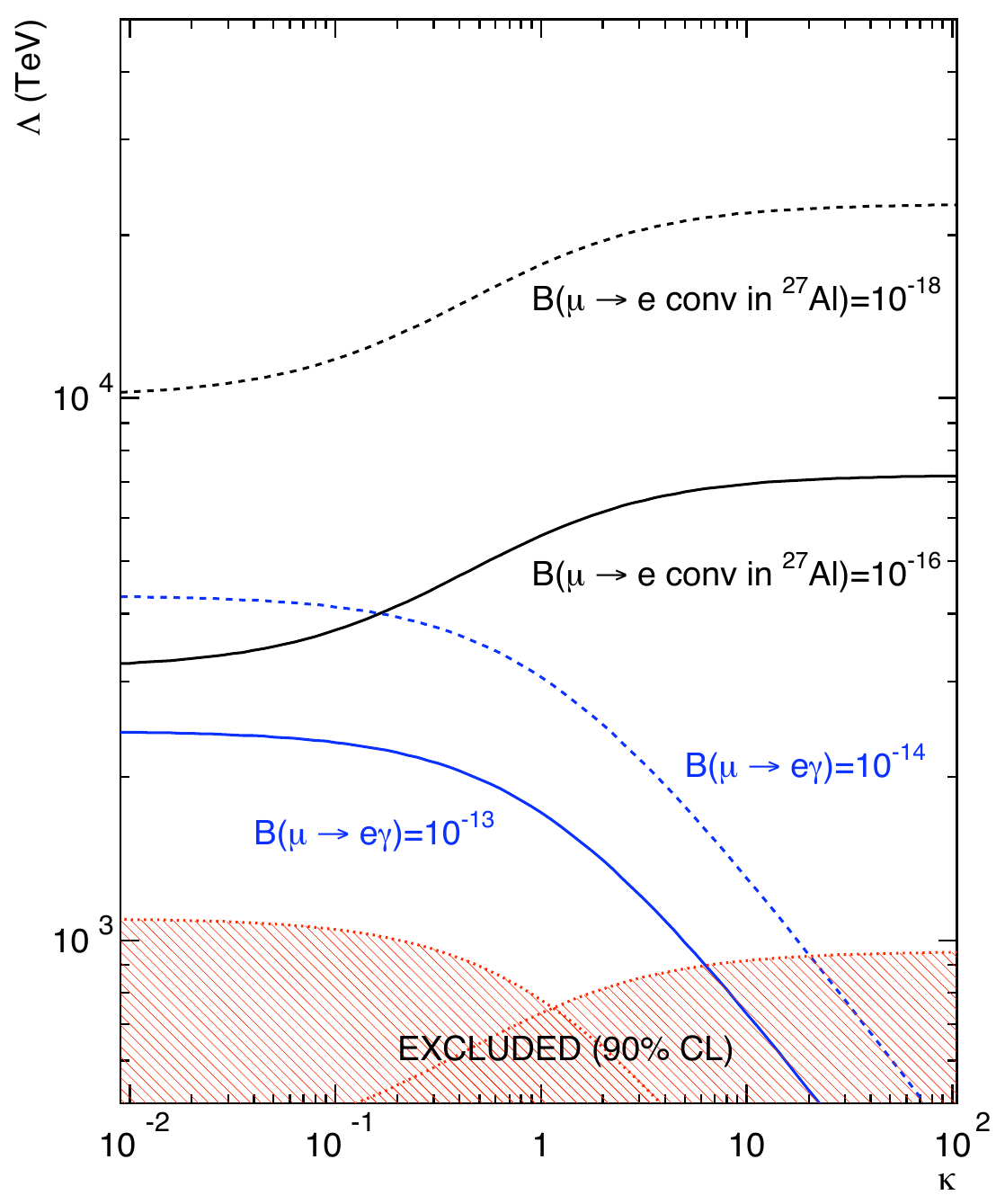}
  \caption{{Sensitivity of a $\mu\to e$ conversion in $^{27}$Al experiment that can probe a normalized capture 
rate of $10^{-16}$ and $10^{-18}$, and of a $\mu \to e \gamma$ search that is sensitive to a branching ratio of $10^{-13}$ and 
$10^{-14}$, to the new physics scale $\Lambda$ as a function of $\kappa$, as defined in the text. Also depicted is the 
currently excluded region of this parameter space. From \cite{deGouvea:2013zba}.
}}
\label{fig:cl:p7}
\end{minipage}
\hspace{0.3cm}
\begin{minipage}[b]{0.48\linewidth}
\centering
    \includegraphics[width=\linewidth]{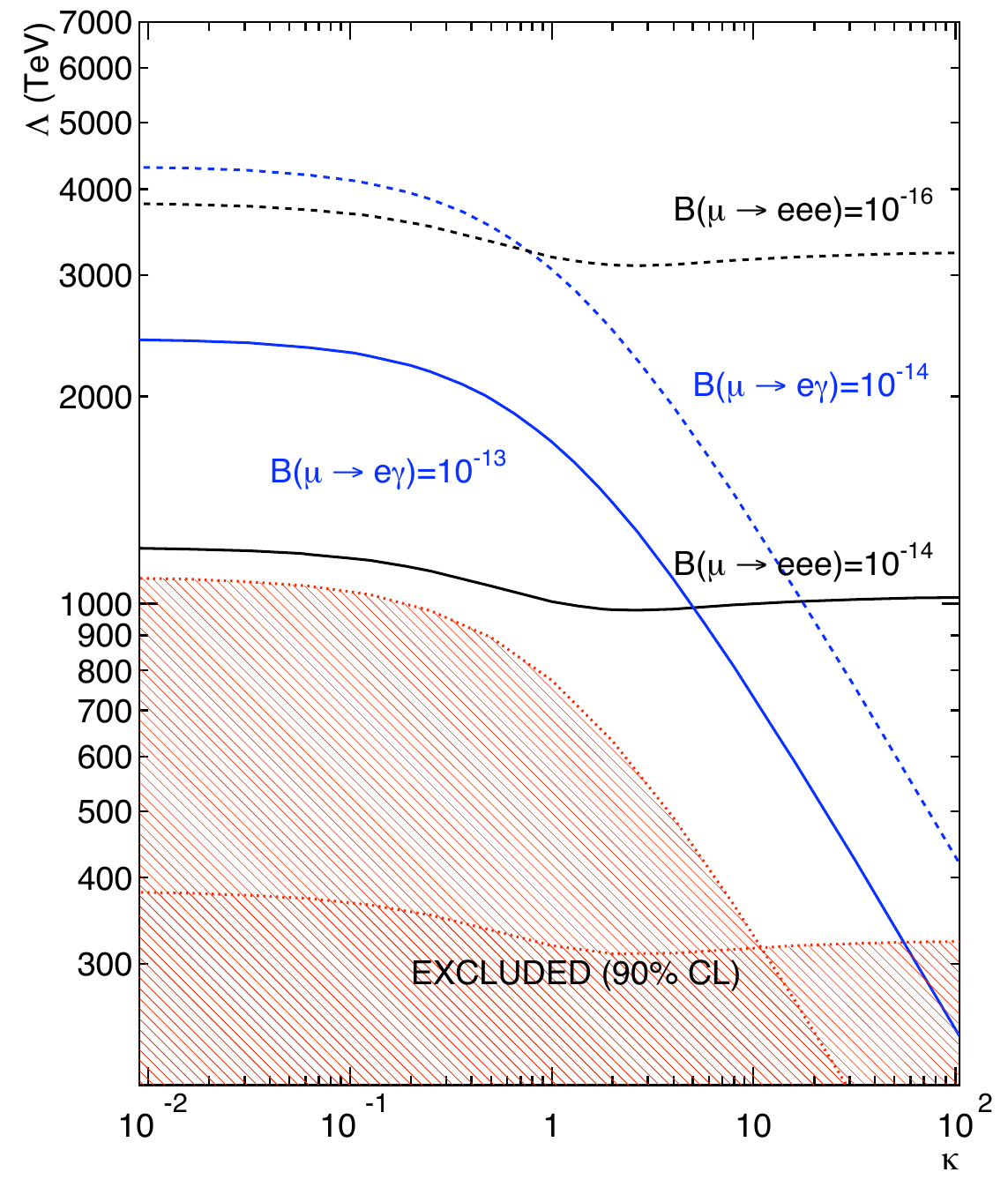}
  \caption{{Sensitivity of a $\mu \to eee$ experiment that is sensitive to branching ratios $10^{-14}$ and 
$10^{-16}$, and of a $\mu \to e \gamma$ search that is sensitive to a branching ratio of $10^{-13}$ and $10^{-14}$, to the new 
physics scale $\Lambda$ as a function of $\kappa$, as defined in the text.  Also depicted is the
currently excluded region of this parameter space. From \cite{deGouvea:2013zba}.
}}
  \label{fig:cl:p8}
\end{minipage}
\end{figure}

Over the next decade,
gains of up to five orders of magnitude are feasible in
muon-to-electron conversion and in the $\mu \to 3 e$
searches, while gains of at least two orders of magnitude
are possible in $\mu \to e\gamma$ and $\tau \to 3\ell$ decay and more than one order of magnitude in $\tau \to \ell\gamma$ CLFV
searches.  
The relative sensitivity to new physics between these processes 
depends on the type of physics amplitude responsible for lepton flavor violation. 
The four-fermion operators that mediate these decays or conversions can be characterized by two parameters: $\Lambda$, which sets the mass scale of the four fermion amplitude, and $\kappa$, which governs the ratio of the four fermion amplitude and the dipole amplitude. $\Lambda$ depends on both the mass and coupling strength of new particles that
may mediate $mu$ to e transitions. 
For $\kappa \ll 1$, the dipole-type operator dominates CLFV phenomena, while for $\kappa \gg 1$, the four-fermion
operators are dominant. Figures~\ref{fig:cl:p7} and \ref{fig:cl:p8}  show these relationships and the capability of new experimental searches which can extend our knowledge quite dramatically in the next decade.

The pattern of violation that emerges yields quite specific information about new physics in the lepton sector. Existing searches already place strong constraints on
many models of physics beyond the Standard Model; the contemplated improvements increase these constraints significantly, covering substantial regions of the parameter space of many new physics models.
These improvements are important regardless of the outcome of new particle searches of the
LHC.  The next generation of CLFV searches is an essential
component of the particle physics road map going forward.  If the LHC finds new
physics, then CLFV searches will confront the lepton sector in ways
that are not possible at the LHC, while if the LHC uncovers no sign of
new physics, CLFV may provide the path to discovery.

In general, muon measurements have the best
sensitivity over the largest range of the parameter space of many new
physics models. There are, however, models
in which  rare tau decays could provide the discovery
channel. Tau flavor violation searches will have their sensitivity extended by around an order of magnitude at new $e^+e^-$ flavor factories. Polarized electron beams can provide an additional gain in sensitivity.  All feasible
CLFV searches should be conducted, since the best discovery
channel is model-dependent and the model of new physics is not yet known.  Should a
signal be observed in any channel, searches and measurements in as
many CLFV channels as possible will be crucial in determining the nature
of the underlying physics, since correlations between the rates
expected in different channels provide a powerful discriminator between
physics models.

\subsection{Flavor-conserving processes}

Flavor-conserving processes have played an important role in defining and understanding the boundaries of the SM.  The measurement of the muon $g-2$ value confirmed the muon was a Dirac particle.  The muon lifetime defines the strength of the weak force, $G_F$.  Low-energy parity-violating scattering experiments confirm the running of the weak mixing angle and provide the best measurements of its value away from the $Z$ pole.  Limits on the electron electric dipole moment (EDM) provide some of the strongest constraints on new sources of CP violation.  The experimental uncertainties on many of these values are at the point where they are sensitive to radiative corrections from TeV scale particles, making them an important complementary probe to new physics searches at the LHC.  The lack of significant discrepancies from SM expectations in flavor-changing processes may indicate a flavor-diagonal or minimal-flavor-violating structure for any new physics at the TeV scale.  This makes the push for improved precision in these flavor-diagonal observables all the more important.

A compelling program of experiments that can be carried out this decade using facilities currently under construction has been assembled.  
These include electron scattering experiments at the JLab 12 GeV upgrade, muon $g-2$ and EDM experiments at the Fermilab muon 
campus and at J-PARC, and tau $g-2$ and EDM measurements at Belle II.  These have been described in detail in the reports associated with the Fundamental Physics at the Intensity Frontier report \cite{Hewett:2012ns}.

As part of the Snowmass process, several initiatives have come to light that have the ability to enhance the currently planned program and also take steps towards the next generation of experiments.  These include major advances in lattice gauge theory that should aid in the interpretation of the muon $g-2$ value.  New initiatives include next-generation measurements of Michel parameters, precise measurements of the weak mixing angle using deep inelastic scattering and an electron-ion collider, and a program for directly measuring fundamental particle EDMs using storage rings.

The muon anomalous magnetic moment $a_\mu$ provides one of the most precise tests of the SM and places important constraints on its extension.
With new experiments planned at Fermilab (E989, also known as Muon $g-2$) and J-PARC (E34) that aim to improve on the current 0.54 ppm measurement at BNL by at least a factor of four, $a_\mu$ will continue to play a central
role in particle physics for the foreseeable future.
In order to leverage the improved precision on $g-2$ from the new experiments, the theoretical uncertainty on the SM prediction must be 
improved and eventually brought to a comparable level of
precision.

The largest sources of uncertainty in the SM calculation are from the non-perturbative hadronic contributions.
The hadronic vacuum polarization (HVP) contribution to the muon anomaly, $a_{\mu}(\rm HVP)$, has been obtained to a precision of 0.6\% using experimental measurements of $e^{+}e^{-}\to\rm hadrons$ and $\tau\to\rm hadrons$.  There is tension between the $e^+e^-$ data from different experiments and between the values of $a_\mu$(HPV) extracted from $e^+e^-$ and $\tau$ data. 
A direct lattice-QCD calculation of the hadronic vacuum polarization with $\sim 1\%$ precision may help shed
light on the apparent discrepancy between $e^{+}e^{-}$ and $\tau$ data;
ultimately a lattice-QCD calculation of $a_{\mu}(\rm HVP)$ with sub-percent precision can circumvent these
concerns.  The HVP contribution to the muon anomalous magnetic moment has been computed in lattice QCD by several groups, as discussed in the Computing Frontier Snowmass report. Statistical errors on lattice calculations of $a_{\mu}(\rm HVP)$ are currently
at about the 3--5\% level, but important systematic errors remain. Anticipated increases in computing resources will enable simulations directly at the physical quark masses with large volumes, and brute-force calculations of quark-disconnected diagrams, thereby eliminating important systematic errors.

Unlike the case for the HVP, the hadronic light-by-light (HLbL) contribution to the muon anomaly cannot be extracted from experiment. Present estimates of this contribution rely on models, and report errors estimated to be in the 25--40\% range.
Therefore an \emph{ab initio} calculation of $a_\mu({\rm HLbL})$ is the highest theoretical priority for $(g-2)$.
A promising strategy to calculate $a_\mu({\rm HLbL})$ uses lattice QCD plus lattice QED, where
the muon and photons are treated nonperturbatively along with the quarks and
gluons.
First results using this approach for the single quark-loop part of the HLbL contribution
have been reported recently.
Much effort is still needed to reduce statistical errors which remain mostly uncontrolled.
In order to bring the error on the HLbL contribution to the level where it is consistent with the projected experimental uncertainty
on the time scale of the Muon $g-2$ experiment, one must reduce the error on $a_\mu({\rm HLbL})$ to
approximately 15\% or better.
Assuming this accuracy, a reduction of the HVP error by a factor of 2, and the expected reduction in
experimental errors, then the present central value would lie 7--8 $\sigma$ from the SM prediction.

Improved measurements of muon decay parameters have the potential to probe
new physics.  While some terms of the effective Lagrangian of muon
decay are tightly bound by the neutrino mass scale if naturalness is
assumed for both Dirac and
Majorana neutrinos, most of the allowed
effective couplings are not constrained by these considerations,
and their best limits come from experiments.

Moreover, $g^{S}_{RL}$ and $g^{V}_{RL}$, the scalar and vector couplings of left-handed to 
right-handed particles, are only constrained by limits on the Majorana nature of neutrinos.  
If a deviation from the SM is found above these constraints that assume a Majorana neutrino,
and if it is possible to establish that the deviation is due to one of these
couplings, it would suggest that neutrinos are Dirac particles.  This
is a unique probe for the existence of Dirac neutrinos, as other known experiments positively test for the
Majorana nature of neutrinos.  The large and versatile muon fluxes available in the next generation 
sources such as Project X make interesting searches possible. Proponents are considering a 
TWIST-like experiment at Project X to perform these searches.

Plans for an electron-ion collider also open several new avenues for the charged lepton program.  In particular, 
the beam polarization makes it possible to construct parity-violating observables in deep inelastic scattering.  This enables 
a measurement of the weak mixing angle over a broad range of momentum transfer.  The 4$\pi$ nature of the detector
 also allows it to cover a large region of phase space, removing limitations associated with fixed target configurations. 
  It is projected that with 200 fb$^{-1}$ at center-of-mass energy of 145 GeV, precise measurements of the weak mixing angle can be 
  made between 10 GeV and the $Z$-boson pole, a region where the value of the mixing angle 
is changing rapidly and has yet to be probed experimentally.

EDMs also play a quintessential role in new physics
searches, particularly in trying to find new sources of CP violation required for electroweak baryogenesis.  
Future experiments may bring us to the point where the lack of evidence for CP violation in EDMs and in quark 
flavor experiments, plus the lack of evidence for new scalar particles at the LHC, may allow us to rule out the electroweak 
phase transition as the source of the matter-antimatter asymmetry in the universe.  This would be a major discovery.

Through the Snowmass process, the group has investigated EDM measurements of charged leptons.
The tau EDM can be probed through triple-vector product correlations in $e^+e^-$ collisions at Belle II.  
The muon EDM can be probed in $g-2$ experiments since the relativistic muons see a motional electric 
field in the rest frame that tilts the precession plane and leads to an asymmetry in the positron decay angle.  
The electron EDM is extracted from limits on EDMs in atoms or molecules.  In these systems, several sources
 of EDMs are possible including the electron, the nucleons, nucleon-nucleon interactions, or  
 electron-nucleon interactions.  Model-independent extractions of the electron EDM are complicated and require multiple systems.

It is possible to make a direct measurement of the electron EDM using a storage ring, analogous to the way $g-2$ is extracted for the muon.  
The key is to use an electrostatic storage ring so that any spin precession can be attributed to an EDM.  Stray radial magnetic fields 
would lead to a false signal.  The effects of such fields can be measured and controlled by using counter-rotating beams. Several years of 
R$\&$D have already been invested into this technique for searching for a proton, deuteron, or muon EDM.  Fortuitously, the ratio
 of the $g-2$ value to the mass of the electron is very similar to this ratio for the proton.  Many systematic effects scale with this ratio so 
 that many studies already performed for the proton can be used for the electron.  Conversely, performing a storage ring EDM 
 experiment on the electron would be an excellent test bed for a (more expensive) proton EDM experiment.  Studies of the
sensitivities of such experiments are underway.
 
 Also fortuitous for the electron is that its magic momentum, the momentum at which the motional magnetic precession 
 cancels, is 15 MeV, requiring a relatively small and inexpensive storage ring.  The technology for the polarized source,
  electrostatic magnets, and beam position monitors all seem to be available.  Concepts for the polarimeter are still being 
  developed and are expected to be the limiting factor in the ultimate sensitivity of the experiment.  A polarimeter with high 
  analyzing power would most likely lead to a sensitivity comparable with model-independent extractions of the electron 
  EDM from atoms and molecules.

%% file: execsumm.tex
\subsection{Overview and summary}
\label{sec:sum-qf}






Flavor-physics experiments probe
very high mass scales, beyond those directly accessible in collider experiments, 
because quantum effects allow heavy
virtual particles to modify the
results of precision measurements.
Even as the LHC embarks on probing the TeV
scale, the ongoing and planned precision flavor-physics experiments are
sensitive to interactions beyond the SM at mass scales which are
higher by several orders of magnitude.
These experiments will provide essential
constraints and complementary information on the structure of models put forth
to explain any discoveries at the LHC, and they have the potential to reveal new
physics that is inaccessible to the LHC.

Throughout the history of particle physics, discoveries made in studies of rare
processes have led to a new and deeper understanding of nature.  A classic example
is beta decay, which foretold the electroweak mass scale and the ultimate
observation of the $W$ boson.  Results from kaon decay experiments were crucial
for the development of the Standard Model;
e.g., the discovery of CP violation in
$K_L^0 \to \pi^+ \pi^-$ decay ultimately pointed toward the three-generation CKM
model.  
Precision measurements of
time-dependent CP-violating asymmetries in $B$-meson decays in the \babar\ and
Belle experiments 
established the CKM phase as the dominant source of CP violation
observed to-date in flavor-changing processes --- leading to the 2008
Nobel Prize for Kobayashi and Maskawa.  

In the past decade our understanding of flavor physics has improved
significantly due to 
\babar, Belle, CLEO, the Tevatron experiments, and most recently LHCb.  
They provided many stringent tests by precisely measuring numerous
CP-violating and CP-conserving quantities.  
The consistency of the measurements with the SM predictions and their 
agreement with CP violation in $K^0$--$\K0bar$ mixing, $\epsilon_K$, 
strengthened the ``new physics
flavor problem."
The new physics flavor puzzle is the question of
why, and in what way, the flavor structure of the new physics is non-generic.
Flavor physics, in particular measurements of meson mixing and CP violation, 
puts severe lower bounds on the scale of new physics,
$\Lambda$.  For some of the most important four-quark operators contributing to
the mixing of the neutral $K$, $D$, $B$, and $B_s$ mesons, the bounds on the
coefficients, $C/\Lambda^2$, 
are at the scale $\Lambda \sim (10^2-10^5)$\,TeV if $C \sim 1$. 
Conversely, for $\Lambda
\sim 1$\,TeV, 
the coefficients 
must be extremely small.
Therefore, there is a
tension between the relatively low (TeV) scale
required to stabilize the electroweak scale and the high scale that is
seemingly required to suppress contributions of physics beyond the SM to flavor-changing processes. 
This problem arises because the SM flavor structure is very special, containing
small mixing angles, and because of additional strong suppressions of
flavor-changing neutral-current (FCNC) processes.  Any TeV-scale new physics
must preserve these features.

The motivation for a broad program of precision flavor physics measurements has
gotten even stronger in light of the first LHC run.
With the discovery of a new particle whose properties are similar to the 
SM Higgs boson, but no sign of other high-mass states, the LHC has begun to 
test naturalness as a guiding principle. 
If the electroweak scale is unnatural, we have
little information on the next energy scale to explore. 
If the
electroweak symmetry breaking scale is stabilized by a natural mechanism, new
particles should be found at the LHC.  
They would provide a novel probe of the flavor sector, and flavor physics
and the LHC data would provide complementary information.  Their combined study
is our best chance to learn more about the origin of both electroweak and flavor
symmetry breaking. 


\subsection{Kaons}
\nopagebreak

Kaon decays have played a pivotal role in shaping the SM,
and they continue to have high impact in constraining the flavor sector of
possible extensions of the SM.
In the area of kaon decays, the FCNC modes mediated by the quark-level processes
$s \to d(\gamma, \ell^+ \ell^-, \nu \bar \nu)$ play a pivotal role.  These include
the four theoretically
cleanest modes
$K^+ \to \pi^+ \nu \overline{\nu}$,
$K_L \to \pi^0 \nu \overline{\nu}$, 
$K_L \to \pi^0  e^+ e^-$, and
$K_L \to \pi^0  \mu^+ \mu^-$.  
Because of the peculiar suppression of the SM amplitude (where the top-quark loop is 
CKM-suppressed by $|V_{td}V_{ts}| \sim \lambda^5$, where $\lambda$ is the Wolfenstein parameter
$\lambda\sim0.2$) which is not present 
in SM extensions, kaon FCNC modes offer a unique window into the flavor
structure of new physics.  This argument by itself provides a strong
and model-independent motivation to study these modes in the LHC era.
Rare kaon decays can elucidate the flavor structure of
SM extensions, information that is in general not accessible from high-energy
colliders.

Discovery potential depends on the precision of the SM prediction for these kaon decays, 
the level of constraints from other observables, and how well
we can measure their branching fractions.  In the modes $K^+ \to \pi^+ \nu
\overline{\nu}$ and $K_L \to \pi^0 \nu \overline{\nu}$,  the intrinsic
theoretical uncertainty is a small fraction of the total, which is currently
dominated by the uncertainty in CKM parameters.  It is expected that in the next
decade progress in lattice QCD and in $B$ meson measurements  from LHCb and
Belle~II will reduce the theory uncertainty on both $K \to \pi \nu \overline{\nu}$
modes to the 5\% level.  

\begin{figure}[t!] 
\centering
\includegraphics[width=0.45\textwidth]{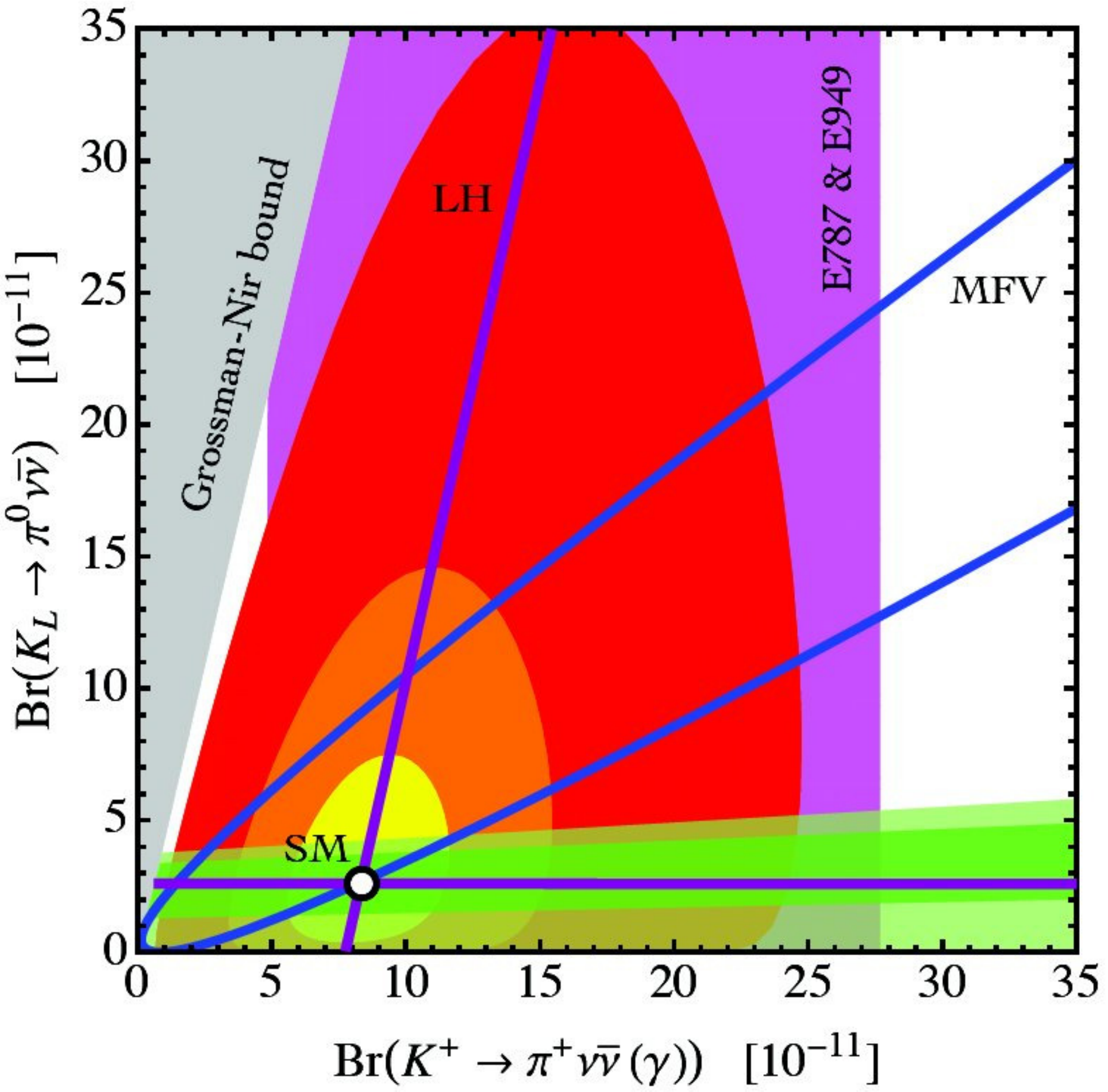}
\caption{\label{fig:constraints} Predictions for the $K \to \pi \nu \bar \nu$
branching ratios assuming dominance of $Z$-penguin operators, for
different choices of the effective new physics couplings  $C_{\rm NP}$. 
The SM point is indicated by a white dot with black border.   The yellow,
orange, and red shaded contours correspond to $|C_{\rm NP}|
\leq \left \{0.5, 1, 2 \right \} |\lambda_t  \hspace{0.5mm}  C_{\rm SM}|$,  the
magenta band indicates the 68\% confidence level~(CL) constraint on ${\cal B}
(K^+ \to \pi^+ \nu \bar \nu \hspace{0.5mm} (\gamma))$ from experiment, and the
gray area is theoretically inaccessible.  The blue parabola represents the
subspace accessible to MFV models.  The purple straight lines represent the
subspace accessible in models that have only left-handed currents, due to the constraint
from $\epsilon_K$.  The green band shows the region accessible taking into
account the constraint from $\epsilon^\prime_K/\epsilon_K$; more general models
do not have this constraint.}
\end{figure}

Rare kaon decays have been extensively studied within well-motivated
extensions of the SM (see Fig.~\ref{fig:constraints}), such as models with supersymmetry
and warped extra dimensions.  In all cases, deviations
from the SM can be sizable, and the
correlations between various rare $K$ decays (as well as with $B$ decays) 
are essential in discriminating among models.  The
$K \to \pi \nu \overline{\nu}$ modes also probe the existence of
light states that are very weakly coupled to the SM, such as those that appear in various dark
sector models, through the experimental
signature $K \to \pi +$\,(missing energy) and distortions to the pion
spectrum.

Besides the FCNC modes, kaon decays also provide exquisite probes of the
charged-current sector of SM extensions, probing the TeV or higher scales. 
Theoretically, the cleanest probes are (1) the ratio $R_K \equiv \Gamma (K \to e
\nu) / \Gamma (K \to \mu \nu)$, which tests lepton universality, scalar, and
tensor charged-current interactions and (2)~the transverse muon polarization
$P_\mu^T$ in the semi-leptonic decay $K^+ \to \pi^0 \mu^+ \nu_\mu$, which is
sensitive to new sources of CP violation in scalar charged-current operators. 
In both cases there is a clean discovery window provided by the precise SM
theoretical prediction of $R_K$ and by the fact that $P_\mu^T$ is generated in the SM 
only by very small and known final state
interactions. 
Table~\ref{tab:exptKsummary} provides a summary of  SM predictions for these
processes, along with current and projected experimental sensitivities  at
ongoing or planned experiments. 

\begin{table}[t]
\centerline{\begin{tabular}{c|c|c|l}
\hline\hline
{Observable} &
  {SM Theory} &
  {Current Expt.} & 
  {Future Experiments}\\
\hline 
${\cal B}(K^+ \to \pi^+ \nu \overline{\nu})$ 
&$7.81(75)(29)\times 10^{-11}$
& $1.73^{+1.15}_{-1.05} \times 10^{-10}$ 
& $\sim$10\% at  NA62 \\
& & E787/E949 & $\sim$5\% at  ORKA \\
& &  & $\sim$2\%  at ProjectX  \\
\hline
${\cal B}(K^0_L \to \pi^0 \nu \overline{\nu})$ 
& $2.43(39)(6)\times 10^{-11}$
&$<2.6 \times 10^{-8}$  \  E391a  &$ 1^{\rm st}$ observation at  KOTO \\
&   &  & $\sim$5\% at  ProjectX  \\
\hline
${\cal B}(K^0_L \to \pi^0 e^+ e^-)$ 
& $(3.23^{+0.91}_{-0.79})\times 10^{-11}$ 
& $<2.8 \times 10^{-10}$ \ KTeV  & $\sim$10\% at  ProjectX \\
\hline
${\cal B}(K^0_L \to \pi^0 \mu^+ \mu^-)$ 
& $(1.29^{+0.24}_{-0.23})\times 10^{-11}$
& $<3.8 \times 10^{-10}$  \ KTeV  &  $\sim$10\% at ProjectX  \\
\hline
$|P_T|$ 
& $ \sim 10^{-7}$ & $<0.0050$  & $<0.0003$ at  TREK  \\
in $K^+ \to \pi^0 \mu^+ \nu$   &   &  & $<0.0001$ at  ProjectX  \\
\hline 
$ \Gamma(K_{e2})/\Gamma(K_{\mu2})$  
& $2.477 (1)  \times 10^{-5} $
& $2.488(10) \times 10^{-5}$  
 &  $\pm 0.0054 \times 10^{-5}$ at TREK \\
   &  &   (NA62, KLOE)  & $\pm 0.0025 \times 10^{-5}$  at ProjectX  \\ 
\hline
${\cal B}(K^0_L \to \mu^\pm e^\mp)$ 
& $< 10^{-25}$  &$< 4.7 \times 10^{-12}$ & $< 2 \times 10^{-13}$ at  ProjectX  \\
\hline\hline
\end{tabular}}
\caption{The reach of current and proposed experiments for some key
rare kaon decay measurements, compared to SM theory and the
current best experimental results. In the SM predictions for $K \to \pi
\nu \bar{\nu}$ and $K \to \pi \ell^+ \ell^-$   the first error is parametric,
and the second is the intrinsic theoretical uncertainty.}
\label{tab:exptKsummary}
\end{table}

A number of rare kaon decay experiments are in progress in Japan and in Europe.
These include: the NA62 experiment at CERN to measure the  $K^+ \to \pi^+ \nu
\overline{\nu}$ branching fraction;  KOTO at J-PARC, which  expects to reach
below the SM level for $K_L \to \pi^0 \nu \overline{\nu}$;  TREK at J-PARC,
which will search for T-violation in $K^+ \to \pi^0 \mu^+ \overline{\nu}$ decays
but also has a broader program of measurements; and KLOE-2 at the Frascati
laboratory, which will improve measurements of neutral kaon interference, 
tests of CPT and quantum mechanics, and  non-leptonic and radiative $K$ decays.

No kaon experiments are currently underway in the U.S. The proposed ORKA
experiment at Fermilab has the potential to utilize the Main Injector
and other existing infrastructure, along with a well-tested experimental
technique,  to make a precise measurement of the  $K^+ \to \pi^+ \nu
\overline{\nu}$ branching fraction based on 1000 or more events.  ORKA will
build on the proven background rejection of BNL E787/E949 (where
seven $K^+ \to \pi^+ \nu \overline{\nu}$ events were observed).  The new
detector will take advantage of technology improvements, but  to achieve the
goals of ORKA,  the background rejection achieved at BNL is adequate. Since the
Main Injector is already scheduled to run to support NO$\nu$A, ORKA 
provides the opportunity to mount
a world-leading rare
kaon decay  experiment in this decade. 

In the longer term, Project X has the potential to provide unprecedented beam power for
producing kaons, leading to at least an order of magnitude higher kaon fluxes. 
Also, the CW-linac of Project X has the ability to provide a well-controlled bunch structure
that can be exploited in rare $K_L$-decay experiments by use of time-of-flight
to measure the decaying $K_L$ momenta.  This provides valuable kinematic
information that will reduce background in a high-statistics $K_L \to \pi^0 \nu
\overline{\nu}$ measurement.

\subsection{$\mathbf{B}$ and charm physics}
\nopagebreak

\begin{table}[t]
\centerline{\begin{tabular}{c|c|c|c}
\hline\hline
\multirow{2}{*}{Observable} &  \multirow{2}{*}{SM theory} 
  &  Current measurement  &  Belle~II \\
  &    &  (early 2013)  &  ($50\, {\rm ab^{-1}}$) \\
\hline 
$S(B \to \phi K^0)$  &  $0.68$  &  $0.56 \pm 0.17$ & $ \pm 0.03$ \\
$S(B \to \eta^\prime K^0)$  & $0.68$  & $0.59 \pm 0.07$  & $\pm 0.02$ \\
$\alpha$ from $B \to \pi\pi,\,\rho\rho$  &  &  $\pm 5.4^\circ$ &  $\pm 1.5^\circ$ \\
$\gamma$ from $B \to D K$  &  &  $\pm 11^\circ$ &  $\pm 1.5^\circ$ \\
$S(B \to  K_S \pi^0 \gamma)$ & $< 0.05$  & $-0.15 \pm 0.20$ & $\pm 0.03$ \\
$S(B \to  \rho \gamma)$ & $<0.05$  & $-0.83 \pm 0.65$ & $\pm 0.15$  \\
$A_{\rm CP}(B \to  X_{s+d}\,\gamma)$  
  &  $<0.005$ &  $0.06 \pm 0.06$ & $\pm 0.02$  \\
$A_{\rm SL}^d$ & $-5 \times 10^{-4}$ & $-0.0049 \pm 0.0038$  & $\pm 0.001$ \\
\hline
${\cal B}(B \to  \tau \nu)$ 
& $1.1 \times 10^{-4}$   & $(1.64 \pm 0.34) \times 10^{-4}$ & $\pm 0.05 \times 10^{-4}$ \\
${\cal B}(B \to  \mu \nu)$ 
& $4.7 \times 10^{-7}$   & $< 1.0 \times 10^{-6}$ & $\pm 0.2 \times 10^{-7}$ \\
${\cal B}(B \to  X_s \gamma)$ 
& $3.15 \times 10^{-4}$  & $(3.55 \pm 0.26)\times 10^{-4}$ & $\pm 0.13 \times 10^{-4}$  \\
${\cal B}(B \to  K \nu \overline{\nu})$ 
& $3.6 \times 10^{-6}$  & $<1.3 \times 10^{-5}$ & $\pm 1.0 \times 10^{-6}$ \\
${\cal B}(B \to X_s  \ell^+ \ell^- )$  ($1 < q^2 < 6$\,GeV$^2$)
& $1.6 \times 10^{-6}$   & $(4.5 \pm 1.0) \times 10^{-6}$ & $\pm 0.10 \times 10^{-6}$  \\
$ A_{\rm FB}(B^{0}\to K^{*0}\ell^{+}\ell^{-})$ zero crossing & 7\% & 18\% & 5\%  \\ 
$|V^{}_{ub}|$ from $B\to\pi\ell^+\nu$ ($q^2>16$\,GeV$^2$) & 
$9\%\to 2\%$ & 11\% & 2.1\% \\
\hline\hline
\end{tabular}}
\caption{The expected reach of Belle~II with 50~ab$^{-1}$ of data
for various topical $B$ decay measurements. Also 
listed are the SM expectations and the current 
experimental results.  
For ${\cal B}(B \to X_s  \ell^+ \ell^- )$, the quoted 
measurement~\cite{Beringer:1900zz} 
covers the full $q^2$ range.
For $|V^{}_{ub}|$ and the $A_{\rm FB}$ zero crossing, we 
list the fractional errors.}
\label{tab:eeBsummary}
\end{table}

\begin{table}[tb]
\centering
\begin{tabular}{c|c|c|c}   \hline\hline
\multirow{2}{*}{Observable} & Current SM   & Precision  &  LHCb Upgrade \\
	& theory uncertainty  & as of 2013 & (50 fb$^{-1}$) \\
\hline
$2\beta_s(B_{s}\to J/\psi \phi)$ &   $\sim0.003$ & 0.09  & 0.008 \\
$\gamma(B\to D^{(*)}K^{(*)})$ & $<1^\circ$ & $8^\circ$ & $0.9^\circ$  \\
$\gamma(B_s\to D_{s}K)$ & $<1^\circ$ & --- &  $2^\circ$ \\ 
$\beta(B^{0}\to J/\psi K^{0}_S)$ &  small & $0.8^\circ$ & $0.2^\circ$  \\ 
\hline
$2\beta_s^{\rm eff}(B_{s}\to \phi\phi)$ & 0.02 & 1.6  & 0.03 \\
$2\beta_s^{\rm eff}(B_{s}\to K^{*0}\bar{K}^{*0})$  & $<0.02$ & --- & 0.02  \\
$2\beta_s^{\rm eff}(B_{s}\to \phi\gamma)$ & 0.2\% & --- & 0.02 \\ 
$2\beta^{\rm eff}(B^{0}\to \phi K^{0}_{S})$ & 0.02 & 0.17  & 0.05 \\
$A_{\rm SL}^s$ & $0.03\times10^{-3}$ & $6\times10^{-3}$ & $0.25\times10^{-3}$ \\
\hline 
${\cal B}(B_{s}\to \mu^{+}\mu^{-}) $ & 8\%  & 36\%  & 5\% \\
${\cal B}(B^{0}\to \mu^{+}\mu^{-}) / {\cal B}(B_{s}\to \mu^{+}\mu^{-})$
  & 5\%  & --- &  $\sim$35\% \\
$A_{\rm FB}(B^{0}\to K^{*0}\mu^{+}\mu^{-})$ zero crossing & 7\% & 18\% & 2\% \\ 
\hline\hline
\end{tabular}
\caption{Sensitivity of LHCb to key observables. The current sensitivity (based
on 1--3\,fb$^{-1}$, depending on the measurement) is compared to that achievable
with 50 fb$^{-1}$ by the upgraded experiment.}
\label{hadronB_tab_1}
\end{table}

The list of interesting observables in $B$ physics is very long.  Two
experiments are poised to exploit this rich promise over the next decade. 
The LHCb experiment has already delivered striking results, for example the
first measurement of $B_s \to \mu^+ \mu^-$. This experiment will continue
running in its current configuration until 2018, when a major upgrade will be
implemented during the LHC shutdown.  Subsequently, LHCb will be able to
collect $5\, {\rm fb^{-1}}$ per year and should reach $50\, {\rm fb^{-1}}$
around 2030.   In Japan the KEK $B$ factory is being upgraded to SuperKEKB, and
the Belle II detector is being built to run there.  SuperKEKB running is planned
to begin in  2017, and  an integrated luminosity of $50\, {\rm ab^{-1}}$ is
projected by 2023. In both cases these large data sets will make dramatic
improvements in sensitivity to new physics across a broad program of measurements.  Examples include
precision measurements of phases and magnitudes of CKM angles ($\beta_s$,
$\gamma$, $|V_{ub}|$, etc.), CP violation in decays dominated by loop diagrams,
leptonic $B$ decays,  and properties of flavor-changing neutral current decays.  
The ATLAS and CMS experiments at LHC
may be competitive on $B$-decay modes with dimuon final states, such as
$B_{s,d} \to \mu^+ \mu^-$.
Belle~II and LHCb will also have broad programs of charm studies, as well
as bottomonium
spectroscopy.

Tables~\ref{tab:eeBsummary} and \ref{hadronB_tab_1} show the current and projected
experimental precision for a number of important measurements
for Belle~II and LHCb, repectively, along with the SM
theory uncertainties.

Searches for new physics in charm decays are complementary to $K$ and $B$
physics, since in FCNC charm decays and in $D^0$--$\bar D^0$ oscillations
intermediate down-type quarks contribute in the loops. 
CP violation in $D^0$--$\bar D^0$ mixing is especially interesting, 
and Table~\ref{Dmixtable} summarizes the
future prospects, using the usual convention for $D^0$--$\bar D^0$ mixing, where the mixing parameters
satisfy $|q/p| =1$ and $\arg(q/p)=0$ in the absence of CP violation.  The study of many
$D$ decay rates and strong phases between amplitudes is also crucial for the $B$
physics program, for example, the extraction of the CKM phase $\gamma$.
In addition, Belle~II and LHCb, along with the dedicated charm experiments
BESIII in China and PANDA in Germany, will study charmonium spectroscopy,
exploring, for example, the so-called $XYZ$ states.

\begin{table}[tb]
\centerline{\begin{tabular}{c|c|c|c}
\hline\hline
Observable  &  Current status  &  Belle~II ($50\, {\rm ab^{-1}}$) 
  &  LHCb upgrade (50 fb$^{-1}$) \\
\hline
$|q/p|$   & 0.91 $\pm$ 0.17  &  $\pm$0.03  &  $\pm0.03$  \\
${\rm arg}\, (q/p)$  &  ($-10.2 \pm 9.2$)$^{\circ}$
  &  $\pm$1.4$^{\circ}$  &  $\pm$2.0$^{\circ}$  \\
\hline\hline
\end{tabular}}
\caption{Sensitivities of Belle~II and LHCb to CP violation in $D^0$ mixing.}
\label{Dmixtable}
\end{table}

\subsection{Theory}
\nopagebreak

To find a convincing deviation from the SM, a new physics effect has to be
several times larger than the experimental uncertainty of the measurement and
the theoretical uncertainty of the SM prediction.  Two
kinds of theoretical uncertainties can be distinguished: perturbative and nonperturbative. 
Perturbative uncertainties come from the truncation of expansions in small (or
not-so-small) coupling constants, such as $\alpha_s$ at a few GeV scale.  There
are always higher-order terms that have not been computed.  Nonperturbative
effects arise because QCD becomes strongly interacting at low energies, and
these are often the limiting uncertainties.  There are, nevertheless, several
possibilities to get at the fundamental physics in certain cases.

\begin{itemize}\vspace*{-12pt}\itemsep 0pt

\item For some observables the hadronic parameters (mostly) cancel, or can be
extracted from data. For example, it is possible to use 
the measured $K\to\pi\ell\nu$ form factor to
predict $K\to\pi\nu\bar\nu$. 

\item In many cases, CP invariance of the strong interaction implies that the
dominant hadronic physics cancels, or is CKM suppressed.  For example, the unitarity
triangle angle $\beta$ can be measured
from $B\to\psi K_S$, and some other CP asymmetries.

\item In some cases symmetries of the strong interaction that arise
in certain limits, such as the chiral or the heavy quark limit, can establish
that nonperturbative effects are suppressed by small parameters.  It is sometimtes
possible to estimate
or extract them from data (e.g., measuring $|V_{us}|$ and $|V_{cb}|$, inclusive
rates).

\item Lattice QCD is a model-independent method to address nonperturbative
phenomena.  The most precise results to date are for matrix elements involving
at most one hadron in the initial and the final state (allowing, e.g.,
extractions of magnitudes of CKM elements).

\end{itemize}\vspace*{-12pt}
Quark flavor physics also provided crucial motivations for developing and
testing numerous effective theories, which have had a big impact for
understanding strong interaction processes at the LHC.

\begin{table}[t]
\centering
\begin{tabular}{cccccc}
\hline\hline
Quantity   & CKM & Present & 2007 forecast & Present  & 2018  \\ 
& element & exp.\ error &\ lattice error\
& \ lattice error\ &\ lattice error\   \\  
\hline
$f_K/f_\pi$ & $|V_{us}|$ \rule[0mm]{0mm}{4mm} & 0.2\% &0.5\%&
{0.4\%} & 0.15\%  \\ 
$f_+^{K\pi}(0)$ & $|V_{us}|$ \rule[0mm]{0mm}{4mm} & 0.2\% & -- &
{0.4\%} & 0.2\% \\ 
$f_D$ \rule[0mm]{0mm}{4mm} & $|V_{cd}|$ & 4.3\% & 5\% &
2\% & $<1\%$ \\
$f_{D_s}$ \rule[0mm]{0mm}{4mm} & $|V_{cs}|$ & 2.1\% & 5\% &
2\% & $< 1\%$  \\
$D\to\pi\ell\nu$ \rule[0mm]{0mm}{4mm} & $|V_{cd}|$ & 2.6\% & -- &
4.4\% & 2\%  \\
$D\to K\ell\nu$ & $|V_{cs}|$ & 1.1\% \rule[0mm]{0mm}{4mm} & -- &
2.5\% & 1\%  \\ 
$B\to D^{*}\ell\nu$ \rule[0mm]{0mm}{4mm} & $|V_{cb}|$ & 1.3\% & -- &
1.8\% & $<1\%$  \\ 
$B\to \pi\ell\nu$ & $|V_{ub}|$ &4.1\% \rule[0mm]{0mm}{4mm} & -- &
8.7\% & 2\%  \\ 
$f_B$ \rule[0mm]{0mm}{4mm} & $|V_{ub}|$ & 9\% \rule[0mm]{0mm}{4mm} 
& -- & 2.5\% & $< 1\%$ \\
$(f_{B_s}/f_B) \sqrt{B_{B_s}/B_B}$ & $|V_{ts}/V_{td}|$ & 0.4\% & 2--4\% &
4\% & $< 1\%$ \\
$\Delta m_s$ & $|V_{ts}V_{tb}|^2$ & 0.24\% & 7--12\% &
11\% & 5\% \\
$B_K$ \rule[0mm]{0mm}{4mm} & ${\rm Im}(V_{td}^2)$
& 0.5\% &3.5--6\% & 1.3\% & $< 1\%$ \\
\hline\hline
\end{tabular}
\caption{History, status and future of selected lattice-QCD calculations
    needed for the determination of CKM matrix elements.}
\label{tab:error}
\end{table}

In the last five years lattice QCD has matured into a precision tool.
A sample of present errors is collected in Table~\ref{tab:error}.
The lattice community is embarking on
a three-pronged program of future calculations:
(1) make significant improvements in ``standard'' matrix elements
of the type just described in the last bullett above,
leading to better precision for CKM parameters;
(2) calculate results for many additional matrix elements relevant for
searches for new physics; and
(3) extend lattice methods to more challenging matrix elements,
which can both make use of old results and provide important information
for upcoming experiments.
These plans rely crucially on access to high-performance computing,
as well as support for algorithm and software development.
The ultimate aim of lattice-QCD calculations is to reduce errors in
hadronic quantities so that other errors dominate. 
As can be seen from Table~\ref{tab:error},
several kaon matrix elements are approaching this level, while
lattice errors remain dominant in most quantities involving
heavy quarks. 
Forecasts for the expected reductions by 2018 are also shown, based on a
Moore's law increase in computing power, and extrapolations using
existing algorithms.
Conservative assumptions project that exascale performance
($10^{18}$ floating point operations/second) will be achieved by 2022,
and that a further factor of 100 will be available by 2032.  These
represent factors of $10^2$ and $10^4$ over currently available
capability.  While  difficult to forecast, this $10^4$ increase in
capability can be expected to significantly expand the range of
quantities that can be computed using lattice methods. 

Thus, the full exploitation of the experimental program requires 
continued support of theoretical developments, including lattice QCD.

\subsection{Conclusions}
\nopagebreak

Because the sensitivities of currently running and proposed flavor physics
experiments are complementary to other searches, phenomena beyond the SM
may be discovered any time the measurements improve.  This gives a strong reason
to pursue flavor-physics measurements whose experimental sensitivity can improve
substantially, and for which the interpretation of results is not limited by theoretical
uncertainties.

The plans in  Europe and Asia appear to be
set for the remainder of this decade, and the 
experiments there (those already running or under construction)
will define the frontier of quark-flavor physics. These are LHCb and NA62 at
CERN, KLOE2 in Italy, Panda in Germany,  BESIII in China, and Belle~II, KOTO,
and TREK in Japan.  This is a rich program, and fortunately U.S.\ physicists
have some involvement in most of these experiments.  While they all have
important physics goals and capabilities, the scale of LHCb and Belle~II, and
their broad physics menus including both bottom and charm, means that
they will be the flagship experiments in quark-flavor physics. In view of that,
the U.S.\ should try to play a significant role in these experiments.

An outstanding question is 
whether the ORKA experiment will go forward at Fermilab.  It 
received ``Stage~1" approval from  Fermilab in the fall of 2011, but has not been 
integrated into DOE's planned program thus far.  
 ORKA presents an extraordinary opportunity, and  
if carried out in a timely way, it can help the U.S.\ HEP program
establish a leading role at the Intensity Frontier.


In the decade beginning around 2020,  LHCb will be well on
its path toward collecting $50\, {\rm fb}^{-1}$ and Belle~II will be well on
its path toward  $50\, {\rm ab}^{-1}$.   
If the U.S.\ HEP program makes a commitment to providing a
high-intensity proton source for the production of neutrino
beams, it will also have the ability to support the next generation of rare kaon decay experiments. 
In particular, Project X  can deliver more than an order of magnitude
increase in the beam power available for producing kaons compared to any other laboratory
in the world.   Project X can be the leading facility in the world
for rare kaon decay experiments.

%% file: nucleons_summary_short_nofigs.tex
%

\newcommand{\mrm}{\mathrm}
\newcommand{\sub}[1]{\mathrm{\scriptscriptstyle{#1}}}

\newcommand{\lampHg}{\ensuremath{{}^{204}\mathrm{Hg}}}
\newcommand{\magHg}{\ensuremath{{}^{199}\mathrm{Hg}}}

\newcommand{\parafield}{\ensuremath{\uparrow\!\uparrow}}
\newcommand{\aparafield}{\ensuremath{\uparrow\!\downarrow}}

\newcommand{\nedm}{\ensuremath{d_{\sub{n}}}}
\newcommand{\hgedm}{\ensuremath{d_{\sub{Hg}}}}
\newcommand{\ecm}{\ensuremath{e\!\cdot\!\mathrm{cm}}}
\newcommand{\ntwoedm}{\ensuremath{\mathrm{n^2EDM}}}
\newcommand{\fTHz}{\ensuremath{\mathrm{fT/\sqrt{Hz}}}}

\newcommand{\tsups}[1]{\textsuperscript{#1}}
\newcommand{\trinat}{{\scshape Trinat}}
\newcommand{\triumf}{{\scshape Triumf}}
\newcommand{\trex}{T{\small REX}}
\newcommand{\zerotozero}{\mbox{$0^+\!\!\rightarrow 0^+\ $}}
\newcommand{\tamutrap}[0]{{\scshape Tamutrap}}

\subsection{Overview}\label{sec:NNAintro}

Tests of fundamental symmetries with nucleons, nuclei, and atoms have played a vital role in developing and probing the 
SM. The observation of parity violation in the radioactive decay of $^{60}$Co, shortly preceding the observation of parity violation in muon decay, provided the first experimental evidence that the weak interaction does not respect this symmetry, ultimately leading to the SM description of charged weak currents as being purely left-handed. Similarly, the measurements of the parity-violating asymmetry in polarized deep-inelastic electron-deuteron scattering in the 1970's  singled out the SM structure for the neutral weak current from among competing alternatives, well in advance of the discovery of the electroweak gauge bosons at CERN. 
Searches for a permanent electric dipole moment (EDM) of the neutron and $^{199}$Hg atom have placed stringent bounds on 
CP violation in the strong interaction, motivating the idea of the spontaneously-broken Peccei-Quinn symmetry, and the associated axion that remains a viable candidate for the cosmic dark matter. 

Today, this use of nucleons, nuclei, and atoms as \lq\lq laboratories" for the study of fundamental interactions is entering a qualitatively new era. The sensitivity of EDM searches in a variety of systems is poised to improve by two orders of magnitude. Measurements of parity-violating (PV) asymmetries in electron scattering and $\beta$-decay are aiming for improvements in precision by an order of magnitude. Advances in theory at the elementary particle, hadronic, nuclear, and cosmological scales will enable us to exploit the results of these experiments in order to shed new light on outstanding questions in fundamental physics. 

At Snowmass, this working group concentrated on EDM searches as well as precise measurements of PV in electron scattering and 
weak decays of light hadrons, as these topics represent a valuable window to physics beyond the SM. 
If the cosmic matter-antimatter asymmetry was produced during the era of electroweak symmetry-breaking (EWSB) -- roughly 10 picoseconds after the Big Bang -- then the CP-violating interactions needed at that time would leave their footprints in non-vanishing EDMs at a level which could be observed with the next generation of experiments. Conversely, the absence of an EDM observation would imply that the matter-antimatter asymmetry was generated at much earlier times, possibly associated with the same dynamics responsible for the tiny but non-vanishing neutrino masses.  CP-conserving interactions associated with the EWSB era that may hold keys to grand unification or weak scale stability could engender tiny deviations of the PV asymmetries from SM expectations. The pattern of such deviations (or their absence) would provide a unique window into an era in cosmic history.
 
\subsection{Electric dipole moments}
\label{sec:EDM}

At the classical level, the EDM of a particle arises from the spatial separation of opposite charges along the axis of the particle's angular momentum. The existence of an EDM would be a direct signature of the violation of both parity (P) and 
time-reversal symmetry (T). 
It also probes the physics of CP violation which necessarily accompanies T violation under the principle of CPT conservation. EDM measurements conducted in many laboratories around the world, employing a variety of techniques, have made tremendous progress, and all have so far obtained results consistent with vanishing EDMs. For example, in the past six decades, the search sensitivity of the neutron EDM has improved by six orders of magnitude to reach the current upper limit of $2.9 \times 10^{-26} \ecm$ .


As discussed above, CP violation in flavor-changing decays of $K$ and $B$ mesons has been observed. The results so far can be 
explained by the CKM mechanism within the framework of the SM, in which all observed weak interaction CP violation phenomena originates 
from a single complex phase. The SM also contains one as-yet unobserved source of CP-violation associated with the strong interaction. The corresponding strength is governed by a dimensionless parameter called \lq\lq ${\bar\theta}$". While one naively 
expects its value to be of order one, the neutron and $^{199}$Hg atom EDM bounds imply that it is at least ten orders of magnitude smaller, hinting at the existence of the spontaneously broken Peccei-Quinn symmetry.

Additional sources of CP violation are generally anticipated in extensions of the SM. For example, in supersymmetry, the supersymmetric partners of the quarks and gauge bosons naturally allow additional complex phases and induce new CP-violating phenomena.  
CP-violating mechanisms beyond the SM are also called for by the observation that the baryon-to-photon ratio in the Universe is as much as nine orders of magnitude higher than the level that can be accommodated by the SM. A much more significant matter-antimatter asymmetry is likely to have been present in the early Universe and provided the favorable condition for the survival of matter that we observe today.

EDMs provide a powerfully sensitive probe for new CP-violating mechanisms, and are a promising path towards discovering 
new physics beyond the SM. The CKM mechanism in the SM can only generate EDMs of elementary fermions at the 
three- and four-loop level, leading to EDM values many orders of magnitude lower than the current experimental limits. For example, the SM prediction for the neutron EDM is expected to be approximately $10^{-31}$ \ecm, or five orders of magnitude below the current upper limit. Any EDM observed in the foreseeable future would require either CP violation in the strong interaction or physics beyond the SM. Perhaps unsurprisingly, extensions of the SM generally allow a range of EDM values that are within the reach of the on-going experiments.  The negative findings so far
provide invaluable constraints on new theories and the upper limits yield insight on the scale of the next Energy Frontier.
The scientific importance and discovery potential of EDM searches 
are strongly endorsed by both the particle and nuclear physics communities\cite{p5,lrp2007}.

The most sensitive EDM searches have so far been conducted on the neutron, nuclei ($^{199}$Hg), and the electron 
(Table \ref{tableedm}). Experiments in these three categories all compete for the prize of being the first to observe a non-zero EDM. However, they complement each other, as each category is most sensitive to different sources of CP violation. For example, the neutron is more sensitive to the EDMs of its constituent quarks; heavy nuclei are more sensitive to the quark chromo-EDM and other CP-violation mechanisms in the nuclear force. The recently proposed storage ring EDM experiments of the proton and deuteron aim to probe combinations of CP-violating contributions that differ from the neutron EDM.  In the future, if a non-zero EDM is discovered in one particular system, it would still be necessary to measure EDMs in other categories to help resolve the underlying CP-violation mechanisms.

\begin{table}
\centering
\caption{Upper limits on EDMs in three different categories.}
\label{tableedm}
\begin{tabular}{|l|c|c|c|}
\hline
 Category & EDM Limit (\ecm)  & Experiment &  Standard Model value (\ecm)   \\
\hline
 Electron             & $1.0\times10^{-27}$    &   YbF molecules in a beam   &   10$^{-38}$       \\
 Neutron             & $2.9\times10^{-26}$    &   Ultracold neutrons in a bottle   &   10$^{-31}$       \\
 Nucleus              & $3.1\times10^{-29}$    &  $^{199}$Hg atoms in a vapor cell     &   10$^{-33}$      \\
\hline
\end{tabular}
\end{table}

At the Intensity Frontier, new sources for cold neutrons and ultracold neutrons (UCNs) are becoming available. The higher output in neutron flux will enable searches for the neutron EDM at a sensitivity level of 10$^{-28}$ \ecm, or two orders of magnitude below the current bound. These experiments could provide a conclusive test of many baryogenesis scenarios, including those that arise in supersymmetry. 

Neutron EDM searches are currently underway at a variety of laboratories around the world, including the Fundamental Neutron 
Physics Beamline at the Spallation Neutron Source (U.S.), the Institute Laue-Langevin (France), 
the Paul Scherrer Institute (Switzerland), the FRM-II reactor at the Technical University of Munich (Germany), 
and at TRIUMF (Canada). Further improvements in neutral atom EDM searches are in progress at the 
University of Washington ($^{199}$Hg), Argonne National Laboratory and Kernfysisch Versneller Instituut (KVI) in the Netherlands ($^{225}$Ra), 
TRIUMF ($^{221}$Rn), and the Technical University of Munich ($^{129}$Xe). More sensitive probes of the electron EDM may be achieved with polar molecules, including YbF (Sussex) and ThO (Harvard-Yale).

The long-term prospects for EDM searches are equally compelling. Future isotope production facilities such as FRIB 
at Michigan State or Project X will produce prolific amounts of selected isotopes that possess enhanced sensitivities to the EDMs of nuclei or electrons. Neutral atoms such as $^{225}$Ra are particularly attractive, as unique features of nuclear structure associated with \lq\lq octupole deformation" can enhance the size of the effect 
at the elementary particle level. An additional  promising direction entails the development of a storage ring EDM search, 
which could enable the direct measurement of the EDM of a charged particle, such as the proton or electron. 
At present, we may only infer limits on the EDMs of these fermions from neutral atom or molecular EDM searches. In these cases, there exist other possible dynamics, such as T- and P-violating electron-nucleus interactions or nucleon-nucleon interactions, 
that could be responsible for any observation of an EDM.  A direct measurement of these charged particle EDMs would open a new era in the tests of CP violation.

\subsection{Neutral currents}
\label{sec:Neutral}

Experiments employing intense beams of polarized electrons scattering from fixed targets, as well as those exploiting parity-forbidden atomic transitions, have played a vital role in developing and testing the electroweak sector of the SM.
Most recently, a program of parity-violating electron scattering experiments performed at MIT-Bates, Mainz, SLAC and Jefferson Laboratory (JLab)  
have provided the most stringent test to date of the energy dependence of the weak mixing angle $\theta_W$ below the weak scale. 
Similarly powerful measurements have been carried out with beams of atomic cesium, yielding the most precise determination of the nuclear weak charge. 
These achievements have built on the steady improvements in experimental sensitivity since the pioneering measurements 
in the 1970's, together with refinements of the theoretical interpretation. The frontier of this field now promises a compelling capability to probe possible new physics at the TeV scale. 


The Qweak collaboration is conducting the first precision measurement at JLab of the weak charge of the proton, $Q^{p}_W$. 
At leading order in the SM, the weak charge depends on the weak mixing angle via the relation
$Q_W^p=1-4\sin^2\theta_W$. Because $\sin^2\theta_W\approx 1/4$, $Q_W^p$ is unusually sensitive to the value of the weak mixing angle as well as to possible deviations from new physics. 
The measurement will determine the weak charge of the proton to the level of 4.1\%, which 
corresponds to constraints on parity-violating new physics at a mass scale of 2.3 TeV at 95\% C.L. This also allows the\
determination of $\sin^{2}\theta_{W}$ to 0.3\% accuracy, providing a sensitive measurement of the energy-dependence
of the weak mixing angle.  

Looking to the future, initial simulations indicate that it is technically feasible to reach a $\sim$2\% sensitivity
on $Q^{p}_W$ at ultra-low energies at the JLab Free Electron Laser (FEL) accelerator complex. Lower-energy measurements 
have a cleaner theoretical interpretation than the Qweak measurement. 
An experiment has been proposed for the newly funded MESA facility at Mainz to pursue a similar
measurement with 2\% sensitivity to 
$Q_W^p$.  The design and required R\&D will be carried out in the next few years so that the experiment
would be ready to start commissioning when MESA first produces external beams, anticipated for 2017.


Parity violation in electron-electron (M\o ller) scattering
has the potential to search for new flavor-conserving amplitudes as small as
$10^{-3}\times G_F$.  The leading term in the PV M\o ller asymmetry is also suppressed by $1-4\sin^2\theta_W$ at tree level. 
The MOLLER experiment has been proposed at JLab to improve on previous
measurement by more than a factor of 5. The goal is a measurement of
the weak charge of the electron
$Q^e_W$ to a fractional accuracy of 2.3\%. 
This would lead to a determination of the weak mixing
angle at the level of $\sim 0.1$\%, which is comparable to the two best such determinations from measurements in $Z$-boson
decays at LEP at CERN and SLC at SLAC.

The most precise measurements of the weak mixing angle have been made from studies of $Z^0$ decays
at LEP and SLC. With the discovery of the Higgs boson at the LHC, the weak
mixing angle is now predicted (within the context of the SM) 
to better precision than all of these measurements individually as well as their combined average. 
While the world average is consistent with the measured mass of the Higgs boson, 
the scatter in the measurements is disconcerting; it would be very useful to have new measurements such as 
MOLLER with comparable precision.



Atomic physics provides
a complementary and powerful probe of the PV neutral current interaction. Atomic PV scales with the nuclear charge roughly as $Z^{3}$, favoring
experiments with heavy atoms. The current most precise measurement of the weak charge of any system has been obtained with atomic cesium, with a relative precision of better than $0.5\%$ (combined experimental and atomic theory uncertainties). Unlike the proton or electron weak charge, the heavy nucleus weak charge is given by $Z(1-4\sin^2\theta_W)-N$. This requires roughly ten times better precision to probe the same level of new physics in these systems. Ongoing efforts in atomic PV include, but are not limited to: Yb (Berkeley), Fr (FrPNC Collaboration at TRIUMF, and the Ferrara, Legnaro, Pisa, Siena collaboration), Ra$^+$ (KVI), and Dy (Berkeley). 

Further improvements in sensitivity  require experiments utilizing as many atoms as possible. 
The planned energy and current for Project X could produce many orders of magnitude more rare isotopes than in any other facility in the world. 
Given the prospective flux, the possibility of multiuser parallel operation mode would give the Project X program a unique capability for 
atomic PV studies in the future.

\subsection{Weak decays}
\label{sec:Weak}

Weak decays of the pion, neutron, and nuclei provide sensitive tests of the universality of the SM charged current (CC) weak interaction.   Electron-muon universality is the hypothesis
that these charged leptons have identical electroweak interactions, differing only by
their masses and coupling to the Higgs boson. However, there could be additional 
effects, such as non-universal gauge interactions, or scalar or pseudoscalar bosons with
couplings not simply proportional to the lepton masses, that would
violate  universality. A sensitive approach to seeking such new interactions is the study of the
ratio 
$
R^\pi_{e/\mu} \equiv {\Gamma(\pi \rightarrow e \nu (\gamma)) / \Gamma(\pi \rightarrow \mu \nu (\gamma))}
$
which in the SM is predicted to be $1.2351(2)\times 10^{-4}$ with a relative uncertainty of  $\pm$ 0.02\%.  New physics at scales as high as 1000 TeV can be
constrained or conceivably unveiled by improved measurements of this ratio. 
Two experiments are currently underway that aim to determine $R^\pi_{e/\mu}$ to 0.05\%: The PIENU experiment  at TRIUMF and  PEN   at PSI. 

In the case of nucleons and nuclei, tests  of the unitarity of the first row of the CKM matrix provide a powerful probe of physics that may violate lepton-quark universality. The CKM unitarity test requires using input from nuclear and/or neutron $\beta$-decays and leptonic decays of
kaons.
Currently, the best-known superallowed decays are from nuclei rather close to stability,
which are easily produced.  However, future improvements in precision will need
comparable measurements on transitions from nuclei much farther from stability.  Higher-intensity
beams will be required to produce them.
Ongoing neutron decay measurements have set the stage for a number of ambitious experiments under
development or construction which target precisions in the $10^{-4}$ range.

%% file: NLWCP_summary.tex




\subsection{Overview}

The Standard Model of particle physics has achieved remarkable success as a result of several decades of exploration, of constantly pushing the boundaries of our knowledge of theory, experiment, and technology.  However, while the SM provides a theoretically consistent description of all known particles and their interactions (ignoring gravity) up to the Planck scale, it is clearly incomplete as it does not address several pieces of evidence for new physics beyond the SM. 

One particularly powerful piece of evidence for new physics comes from the existence of dark matter (DM).  DM dominates the matter density in our Universe, but very little is known about it.  Its existence provides a strong hint that there may be a dark sector, consisting of particles that do not interact with the known strong, weak, or electromagnetic forces.  Given the intricate structure of the SM, which describes only a subdominant component of the Universe, it would not be too surprising if the dark sector contains a rich structure itself, with dark matter making up only a part of it.  Indeed, many dark sectors could exist, each with its own beautiful structure, distinct particles, and forces. 
These dark sectors may contain new light weakly-coupled particles (NLWCPs), particles well below the weak scale that interact only feebly with ordinary matter.  Such particles could easily have escaped past experimental searches, but a rich experimental program has now been devised to look for several well-motivated possibilities.  

The existence of dark sectors is motivated by bottom-up and top-down theoretical considerations.  These sectors arise in many theoretical extensions to the SM, such as moduli that are present in string theory or new (pseudo-)scalars that appear naturally when symmetries are broken at high energy scales.  Other powerful motivations include the strong CP problem, and various experimental findings, including the discrepancy between the calculated and measured anomalous magnetic moment of the muon and puzzling results from astrophysics. 
Besides gravity, there are a few interactions allowed by SM symmetries (given below) that provide a ``portal" from the SM sector into the dark sector.  These portals include

\begin{center}
\begin{tabular}{rcl}
``Vector'' portal: & Dark photons & $-\frac{\epsilon}{2\cos\theta_W}B_{\mu\nu}F'^{\mu\nu}$ \\
``Axion'' portal: & Pseudoscalars & $ \frac{\partial_{\mu}a}{f_{a}} \overline{\psi}\gamma^{\mu}\gamma^{5}\psi$ \\
``Higgs'' portal: & Dark scalars & $ (\mu S + \lambda S^{2})H^{\dagger}H $ \\
``Neutrino'' portal: & Sterile neutrinos & $ y_N LHN $ \\
\end{tabular}
\end{center}

The Higgs and neutrino portals are best explored at high-energy colliders and neutrino facilities, respectively.  Our focus here will be on the vector and axion portals, which are particularly compelling possibilitites and can be explored with low-cost, high-impact experiments.

\subsection{Axions and axion-like particles}

\begin{figure}[ht]
\centering
\includegraphics[width=0.75\textwidth]{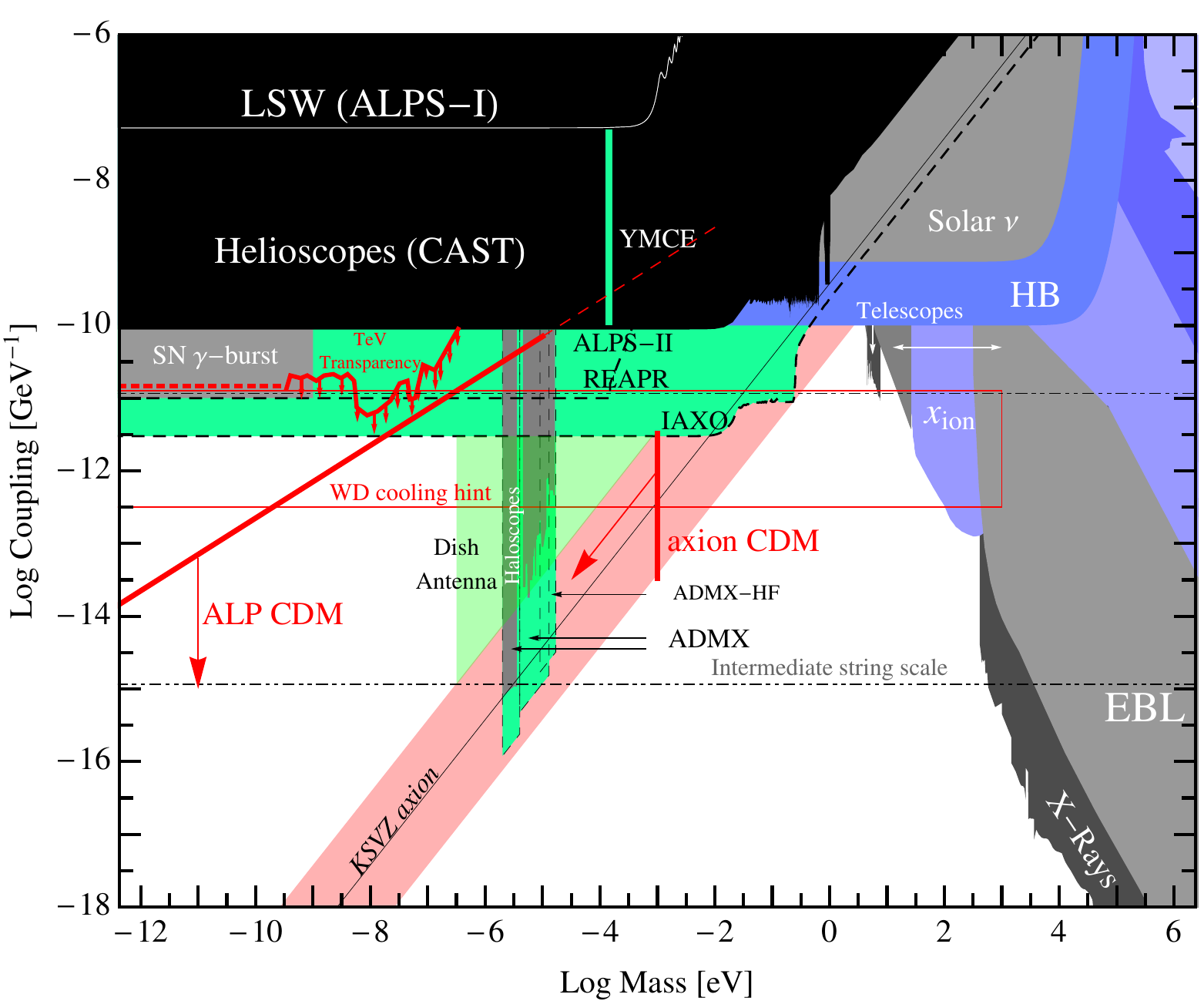} \\
\caption{Constraints and future prospects for axion and axion-like particle searches in the coupling-mass plane.
}\label{fig:ALPs}
\end{figure}

The QCD axion has an origin that lies as the solution to the strong-CP problem of Quantum Chromodynamics (QCD). As discussed 
in the previous section, limits on the QCD vacuum angle parameter from the non-observation of a neutron electric dipole moment give $\overline{\theta} < 10^{-10}$, when any angle between 0 and 2$\pi$ would have been equally likely. The axion was introduced as the Goldstone boson associated with the breaking of the Peccei-Quinn symmetry that makes the value of $\overline{\theta}$ near zero. Interestingly, 
the axion is a natural candidate to contribute to the DM of the universe, with thermal production, a process known as the ``misalignment mechanism," and radiation from cosmic strings or domain walls all possibly contributing to the axion matter density. The QCD axion theory has interrelated parameters describing its mass and couplings to other SM particles.
For instance, the axion mass is given approximately by the following:
\begin{equation}
m_a \sim \frac{\Lambda_{\rm QCD}}{f_a} m_\pi \sim \frac{0.6~{\rm meV}}{f_a / 10^{10}~{\rm GeV}}
\end{equation}
where $f_a$ is the Peccei-Quinn symmetry breaking scale. The coupling to two
photons is often used in experimental searches and is also suppressed by the
high symmetry-breaking scale. The pseudo-scalar nature of the axion gives rise to a coupling dependent on $g_{a \gamma \gamma} \vec{E} \cdot \vec{B}$, where
\begin{equation}
g_{a \gamma \gamma} \sim \frac{\kappa}{f_a}
\end{equation}
and $\kappa$ represents a range of model parameters and includes a factor of $\alpha/4\pi$. Fig.~\ref{fig:ALPs} shows current constraints on axions, where the diagonal band of coupling versus mass represents what is expected for the QCD axion from the cosmological abundance. An axion-like particle (ALP), in principle, can have mass and couplings anywhere in the allowed region since its properties are not necessarily prescribed within the framework of QCD
but are still motivated as one of the portals to the dark sector. Fig.~\ref{fig:ALPs} shows a number of experimental constraints on the parameter space for axions and axion-like particles that highlight the current and near future plans. Also shown are constraints on the coupling that require $g_{a \gamma \gamma} \lesssim 10^{-10}$~GeV$^{-1}$ due to the observation that star cooling is well-described without an axion-like particle cooling mechanism.  We now discuss the various search techniques.

{\bf Microwave cavities:}
The Axion Dark Matter Experiment, ADMX, is an ongoing program to detect possible dark matter axions or ALPs by using a high-Q tuned microwave cavity immersed in an 8T magnetic field. If a dark matter axion enters the cavity and if the energy of the axion matches the cavity frquency, then a possible signal can be detected above the dominant thermal noise. By adjusting tuning rods within the cavity, scans in frequency provide sensitivity over a range of axion masses. The experiment has been approved to continue running with very-low-noise SQUID detectors and with an upgrade to a dilution refrigerator to improve sensitivity. The current plans are to scan over a mass range in the region between approximately $10^{-6}$ and $10^{-5}$~eV and to cover the QCD axion band. Experiments using smaller cavities may be able to probe the next decade of masses up to $10^{-4}$~eV in the QCD axion band.  The ADMX program is complementary to the WIMP dark matter direct detection experiments in the overall portfolio of searching for likely dark matter candidates.

{\bf Light shining through the wall and solar helioscopes:}
One search strategy for axion-like particles is a laser light-shining-through-a-wall (LSW) experiment, where a laser beam is directed into the bore of a dipole magnet. Through the axion-photon-photon coupling, some of the laser photons, in the presence of the magnetic field, may convert into ALPs, which, in turn, are weakly interacting and may propagate through an opaque barrier, the ``wall." The ALPs may be detected if they reconvert into detectable photons in the magnetic field region on the other side of the wall. A world-wide effort excluded a
claim that an ALP had been detected within the
last decade. Experiments at Fermilab, JLab, CERN, and DESY set the limit 
$g_{a \gamma \gamma} \lesssim 10^{-7}$~GeV$^{-1}$. Future efforts include developing a resonantly enhanced axion-photon regeneration LSW experiment where phased-locked cavities on both the generation and regeneration sides of the wall enhance the sensitivity to the coupling to probe the region around $g_{a \gamma \gamma} \lesssim 10^{-11}$~GeV$^{-1}$, where there are astrophysical hints of its relevance. Some of the increased sensitivity arises from utilizing a much longer string of either Tevatron (REAPR) or HERA (ALPS-II) dipole magnets.

If ALPs are produced in the Sun, then an analogous LSW process might occur that would allow for the regeneration of solar axions back into X-ray photons that could be detected by observing the sun opaquely though an accelerator magnet. The CAST helioscope project has used this strategy to probe the $g_{a \gamma \gamma} \lesssim 10^{-10}$~GeV$^{-1}$ region and has extended that search over a broad range of masses including the QCD axion band around 1~eV where astrophsyical constraints from hot dark matter (HDM) also apply. The follow-up concept is the $4^{th}$-generation helioscope international axion X-ray observatory (IAXO) that would require a new custom large toroidal magnet capabable of tracking the Sun. Besides X-ray detectors,
IAXO would make available ports for other technologies that might have sensitivity to more exotic NLWCPs.

\subsection{Dark photons and other dark-sector states}

Extensions of the SM can give rise to extra U(1) gauge symmetries, such as dark-sector analogues of the SM hypercharge U(1)$_{\rm Y}$.  Such a U(1) can give rise to a massive vector boson, often called a dark photon and abbreviated as $A'$, which may mediate interactions among dark-sector particles (including DM) charged under the U(1).  A generic mechanism known as kinetic mixing generates a mixing between the $A'$ and the SM photon at energies below the weak scale. This in turn induces a weak coupling between the $A'$ and electrically charged SM particles, providing a portal into the dark sector.  This coupling allows $A'$s to be produced in charged particle interactions and, if sufficiently massive, decay into pairs of charged particles like $e^+e^-$.  DM interactions with an $A'$ can also produce signals  in DM direct and indirect detection, as well as other types of experiments. For very light $A'$, kinetic mixing allows for $A' \leftrightarrow \gamma$ oscillations, analogous to neutrino oscillations.  

In contrast to the other portals, kinetic mixing and the vector portal generates an interaction that is unsuppressed by a high mass scale.  Simple mechanisms for kinetic mixing lead us to expect $A'$ couplings to electrically charged particles of $\epsilon\/ e$, where $\epsilon \sim 10^{-2}$ to $10^{-5}$ and $m_{A'}\sim \sqrt{\epsilon}\ m_Z \sim {\rm MeV ~to~ GeV}$, making production and detection of $A'$s possible at existing collider and fixed-target facilities. If $A'$ decays into dark-sector particles are kinematically allowed, these ``invisible'' decays may dominate and produce light DM particles, which could in turn interact through $A'$ exchange and be detected. 
While the MeV to GeV mass range is a particularly motivated region to explore, other regions with lower $A'$ masses and lower values of $\epsilon$ are also natural in simple extensions of the SM, and should thus be rigorously pursued as well.  


The existence of dark photons is also motivated by anomalies in particle physics and astrophysics, and from recent hints of DM detection in direct detection experiments.  Observation of an excess of high-energy electrons and positrons in the cosmic-rays, most recently confirmed by AMS-02, re-invigorated the search for an $A'$. The persistent discrepancy between the experimental and SM-predicted values of $g-2$ of the muon could be evidence for a $\sim 10$~MeV $A'$ with $\epsilon \sim 10^{-3}$.  Recent candidate events of $\sim 10$~GeV DM in direct detection experiments are hard to accommodate if the DM has only SM interactions, but are natural in models with a new light mediator.  An $A'$ with mass well below the MeV-scale can itself constitute all the DM in a large region of parameter space.  

It is useful to consider three cases in detail:  

\begin{figure}[!t]
\centering
\vspace*{-5mm}
\includegraphics[width=0.47\textwidth]{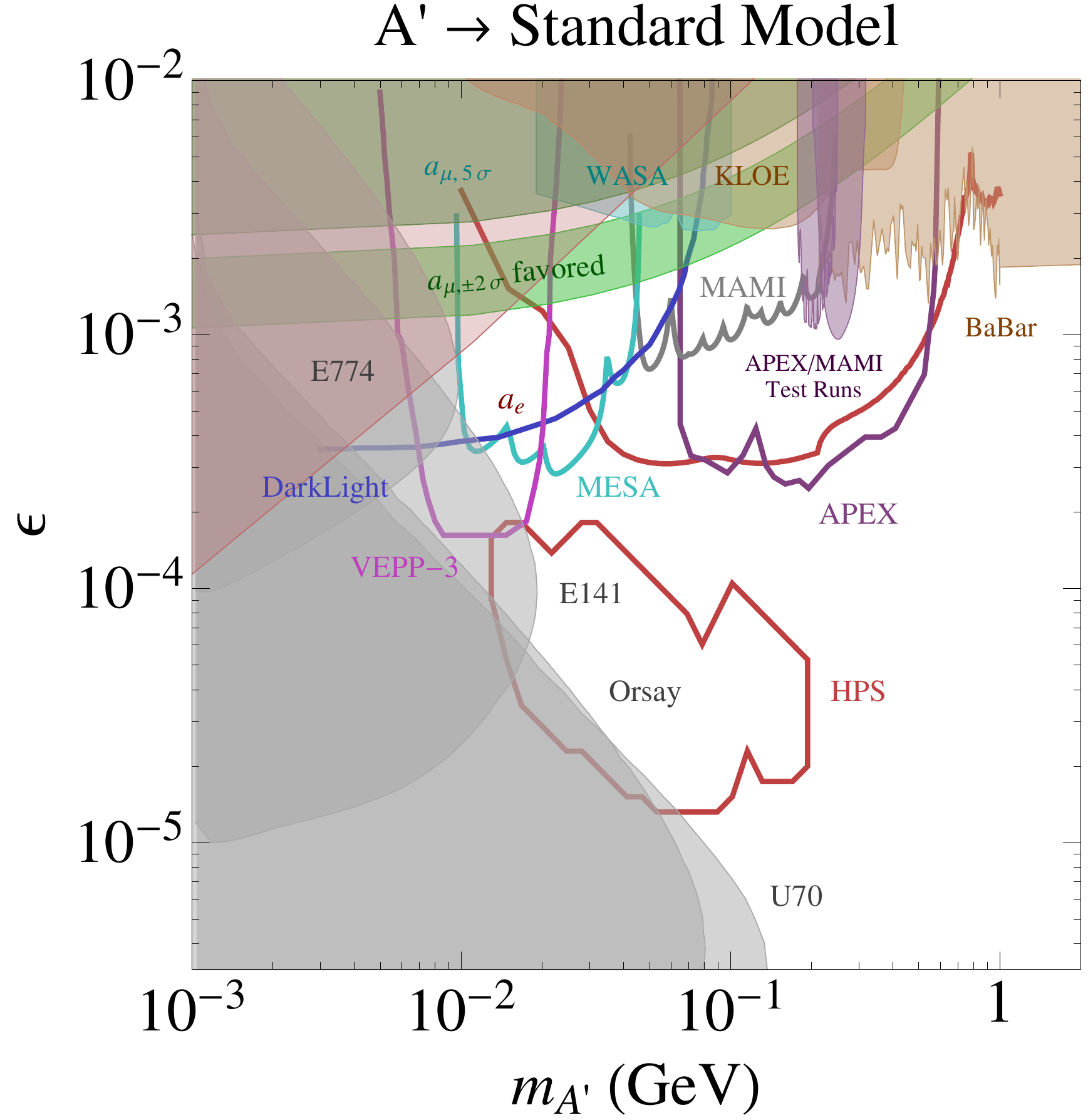} \;\; 
\includegraphics[width=0.47\textwidth]{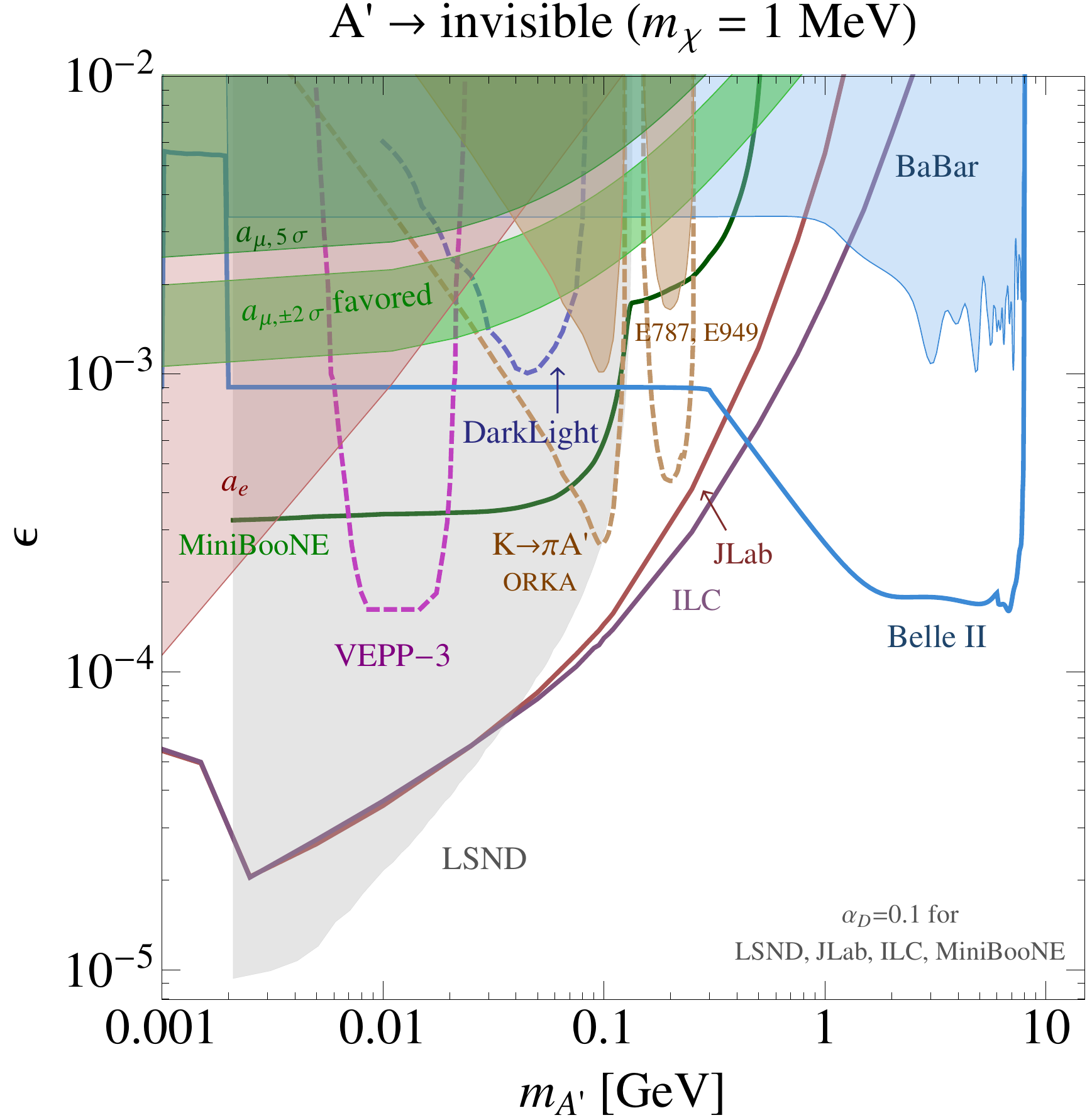}\;\; 
\caption{{\bf Left:} Existing constraints and future opportunities to search for dark photons with masses above 1 MeV decaying visibly to Standard Model states (e.g.~$e^+e^-$).  Opportunities over the next decade include experiments at JLab, Mainz, and VEPP-3, as well as Belle II and others. {\bf Right:}  Existing constraints and future opportunities to search for dark photons with masses above 1 MeV decaying invisibly to light dark-sector states, including sub-GeV DM, \emph{assumed to have a mass of 1~MeV}.  Opportunities over the next decade include proton dumps (e.g.~with MiniBooNE, LSND, NOvA, Project X), electron dumps (e.g.~at SLAC, JLAB, SuperKEKB, ILC), Belle II, ORKA, and electron fixed-target experiments (e.g.~DarkLight and VEPP-3).  Note that existing constraints and prospects of different experiments can change \emph{drastically} for different masses of the dark-sector states, for different A'-to-dark-sector coupling, and for minor changes in the model; this plot is thus not representative of the full parameter space that needs to be explored.  See text and \cite{NLWCP} for more details.  
}
\label{fig:heavy-A'}
\end{figure}

\begin{itemize}
\item[(1)] $m_{A'} \sim $~MeV to GeV with \emph{visible} $A'$ decays: 
A series of re-interpreted beam dump experiments, measurements of $g-2$ of the electron and muon, and astrophysical limits complement recent direct searches at $e^+e^-$ colliders and fixed target electro-production experiments, to constrain the region shown in Fig.~\ref{fig:heavy-A'} (left).  The region for which an $A'$ can explain the muon $g-2$ discrepancy is shown in a green band.  The reach of several fixed-target experiments (APEX, HPS, and DarkLight at JLab and experiments at MAMI and VEPP-3) are delineated with lines.  These are now in the planning phase or under construction and will explore much of the motivated parameter space.  Future searches (not shown) at the very high luminosity $e^+e^-$ colliders Belle~II and KLOE~2 complement these searches, but leave unexplored higher masses and lower couplings.  New ideas for future experiments (not shown) suggest much broader coverage is possible. 


\item[(2)] $m_{A'} \sim $~MeV to GeV with \emph{invisible} $A'$ decays:
Constraints and opportunities to look for invisible decays of the $A'$ are still being actively worked out.  Fig.~\ref{fig:heavy-A'} (right) shows the constraints from the electron and muon $g-2$ (along with the muon $g-2$ preferred region), rare kaon decays, and a reinterpreted BaBar analysis.  Also shown are future opportunities for ORKA, Belle II, DarkLight, and VEPP-3, which use kinematic constraints to identify invisible $A'$ decays. Invisible decays can also be detected by observing the interactions of the hidden sector particles (for example, sub-GeV dark matter) produced in $A'$ decays. Both electron and proton beam dump experiments could produce an $A'$, which could decay directly to these particles. They in turn would escape the dump and could interact downstream in detectors. Also shown are the proposed reach for MiniBooNE to search for sub-GeV dark matter and future opportunities for electron beam-dumps at JLab, SLAC, KEK, MAMI, ILC etc. These experiments will complement searches for sub-GeV DM in direct detection experiments.

\begin{figure}[ht]
\centering
\includegraphics[width=0.8\textwidth]{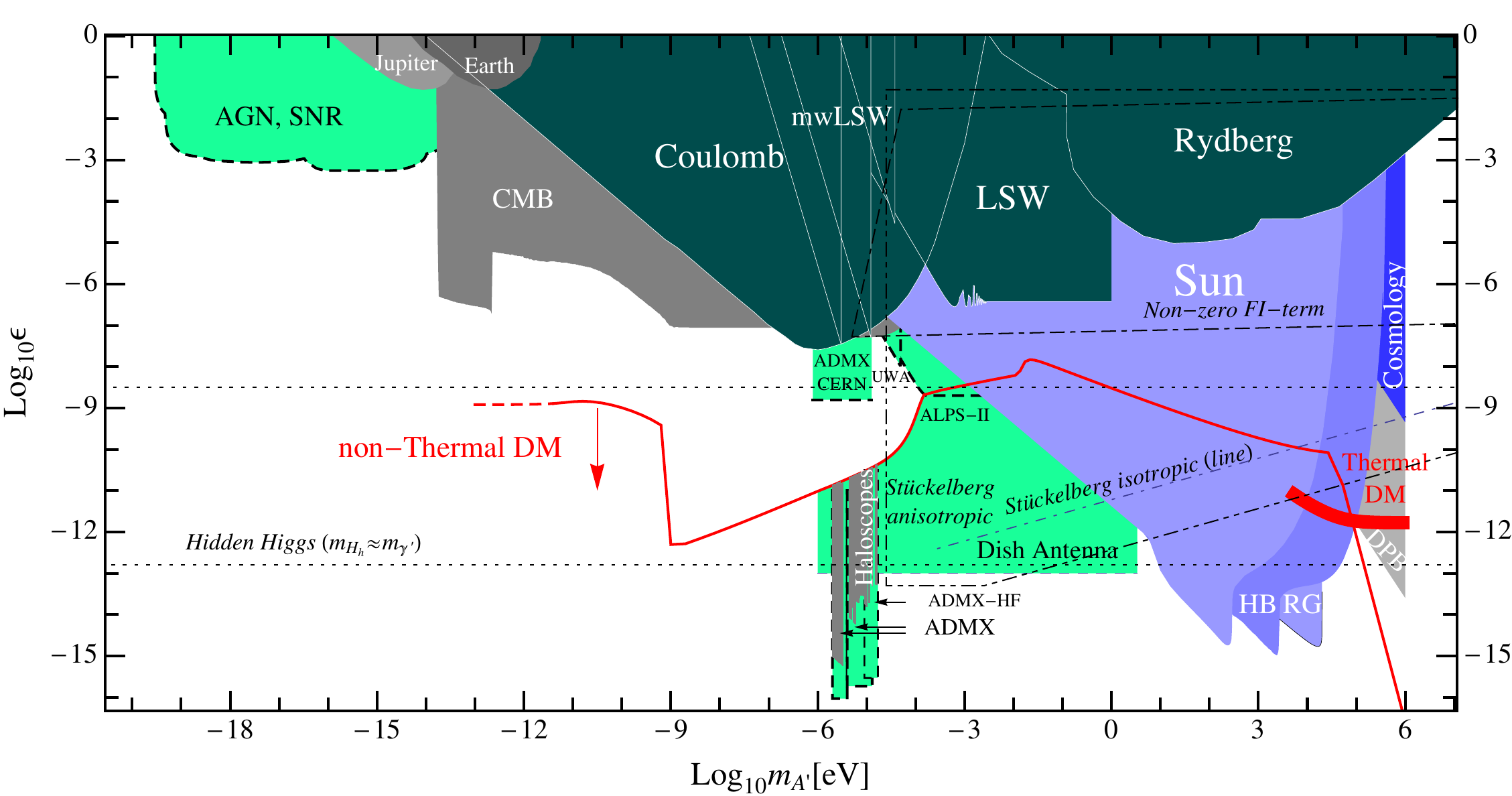}\;\; 
\caption{
Parameter space, constraints, and opportunities for dark photons with masses below 1 MeV. The region shown in light green remains accessible to future experimentation. 
}
\label{fig:light-A'}
\end{figure}
\item[(3)] $m_{A'} <$~MeV: Below an MeV, the $A'$ is unable to decay to $e^+e^-$ and is thus long-lived.  Existing constraints and opportunities for future searches are shown in Fig.~\ref{fig:light-A'}. Constraints arise from stellar cooling, precision measurements of Coulomb's law, and past light-shining-through-walls (LSW) experiments.   A large parameter space (shown in light green) remains accessible to future experimentation, including regions in which the $A'$ itself could constitute all of the DM.  Experiments searching for axions and ALPs can take advantage of $A'\leftrightarrow \gamma$ oscillations and probe low-mass dark photons.   
\end{itemize}

\subsection{Chameleons}

\begin{figure}[htb]
\begin{center}
\includegraphics[width=0.7\textwidth]{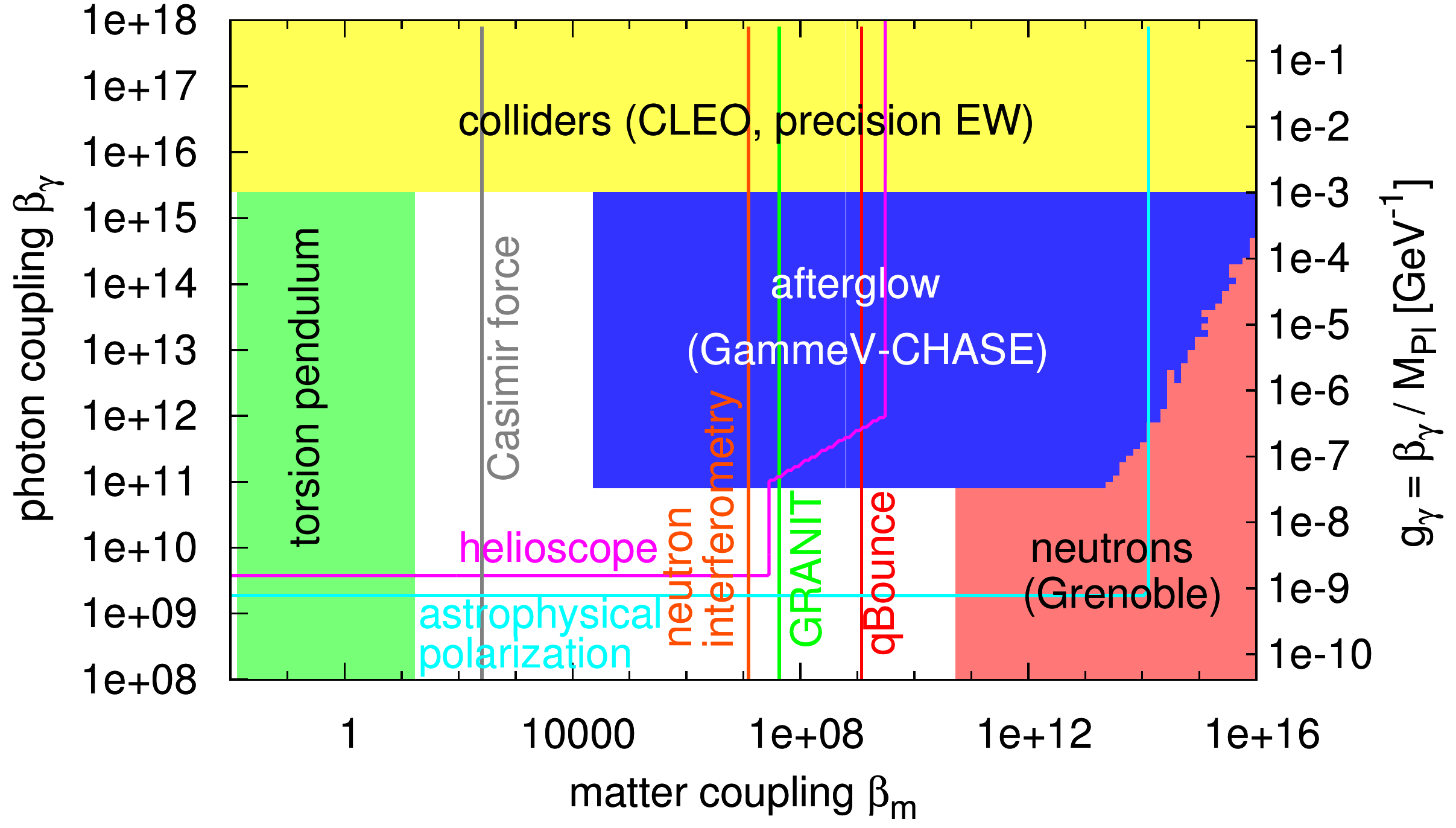}
\caption{Constraints on the matter and photon couplings for a chameleon dark energy model with $V(\phi) = M_\Lambda^4(1+M_\Lambda/\phi)$.  Current constraints are shown as shaded regions, while forecasts are shown as solid lines.  
\label{fig:chameleon_constraints}}
\end{center}
\end{figure}

Chameleon particles are new particles predicted in some theories that have properties that depend on their environment. In particular, the chameleon is nearly massless in regions with low matter density and becomes massive in regions of high matter density. Such particles have properties that could explain a possible scalar field responsible for the accelerated expansion of the universe (dark energy) while avoiding constraints from terrestrial laboratory and solar system fifth force experiments. The chameleon mechanism ($m_{{\rm eff}} \propto \rho$) arises naturally with a scalar coupling to the stress energy tensor within a wide range of possible potentials. An afterglow experiment, GammeV-CHASE, set limits on the chameleon coupling to photons by using a reconfigured LSW apparatus. Other experiments using neutrons and a torsion pendulum also constrain the phase space of allowed parameters. These limits include those from colliders sensitive to this physics at the TeV scale, as shown in Fig.~\ref{fig:chameleon_constraints}.

%% file: benchmarks.tex
As part of the Snowmass study, an investigation of a well-chosen 
set of benchmark scenarios was performed.  These studies provide a quantitative
comparison of experimental capabilities across the various programs of the
Intensity Frontier.  They also provide the opportunity to demonstrate connections
across the Intensity, Energy, and Cosmic Frontiers.  Here, we report on flavor
studies in supersymmetry and warped extra dimensions as well as signatures of neutrino 
models that span the frontiers.

\begin{figure}[ht]
\centering
\includegraphics[width=0.75\textwidth]{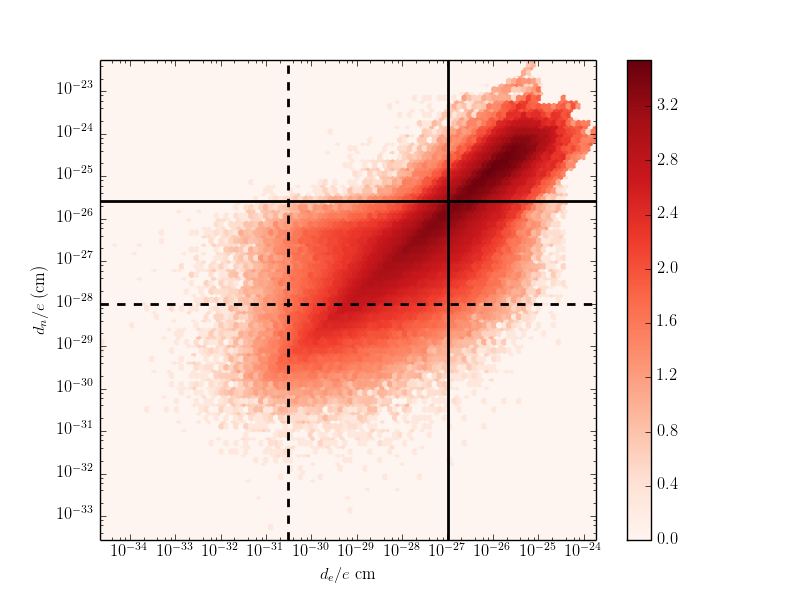} \\
\caption{Values of the neutron EDM versus the electron EDM in the pMSSM including a scan over
allowed phases. The solid (dashed) lines indicate present (future) experimental reach.  This set of
pMSSM models is not expected to be observed at the high-luminosity 14 TeV LHC.  From \cite{Berger:2013zca}.}
\label{fig:pMSSM}
\end{figure}

The Supersymmetry study was based on the phenomenological MSSM (pMSSM)\cite{Berger:2013zca}.  The pMSSM assumes the
most general CP-conserving model with R-parity, minimal flavor violation (MFV), and does not
employ any GUT, supersymmetry-breaking, or any other assumptions about the nature of
physics at high scales.  This set of guiding principles results in a 19-dimensional
supersymmetric parameter space.  This study took the lightest supersymmetry particle to be the
lightest neutralino.  A sample of $\sim 225,000$ such models was generated by
scanning over the parameter space -- where all
supersymmetric particle masses were set to be $< 4$ TeV -- and are
consistent with the global data set on collider, astrophysical, precision electroweak,
and heavy flavor measurements.  This set of pMSSM models was employed for Snowmass studies in all
three physics frontiers.  For Intensity Frontier studies, the dimensionality of the
parameter space was enlarged to include an additional scan over CP-violating phases, and a scan over the parameters
representing an expansion in MFV.  Several observables from
Intensity Frontier measurements, including EDMs and quark flavor and charged lepton processes,  
provide a large discovery window which is
complementary to that of experiments at the other Frontiers.  In particular, pMSSM models
that are not expected to be detected at the 14 TeV, 3 ab$^{-1}$ LHC were found to have
observable signatures in several proposed Intensity Frontier experiments; an example employing
electric dipole moments is
shown in Fig.~\ref{fig:pMSSM}.

\begin{figure}[ht]
\centerline{
\includegraphics[width=0.45\textwidth]{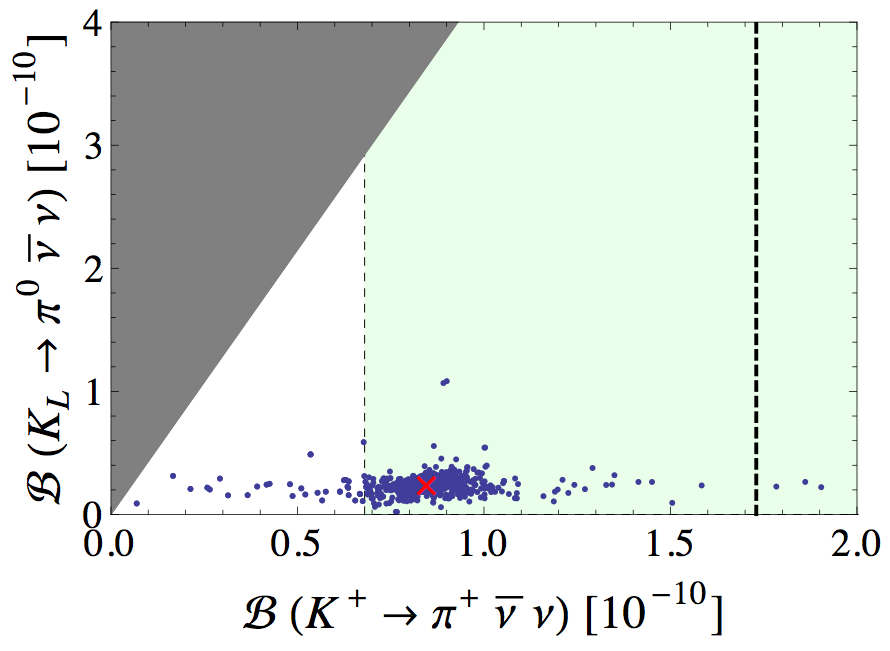} 
\includegraphics[width=0.45\textwidth]{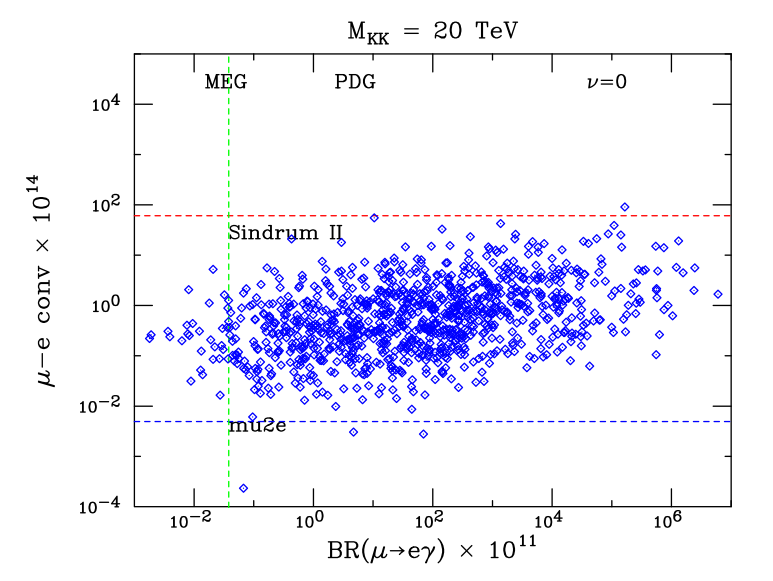}} 
\caption{Branching fractions of charged versus neutral kaon decay modes (the red \lq x' indicates the SM value) (left) and rates for $\mu N\to eN$
versus $\mu\to e\gamma$ (right) in benchmark models of warped extra dimensions.  
Present and proposed future experimental reaches are as indicated.  
From \cite{Agashe:2013kxa}.}
\label{fig:RS}
\end{figure}

Intensity Frontier measurements provide an excellent probe of
theories with warped extra dimensions.  A  Snowmass study\cite{Agashe:2013kxa} focused on Randall-Sundrum models of quark flavor,
which explain the quark mass and mixing hierarchies in terms of spatial geometries, {\it i.e.,} the localization
parameters of the fermion's extra-dimensional wavefunction.  In this scenario, all SM fermions and gauge bosons
reside in the additional dimension, resulting in a Kaluza-Klein tower of particles in our four-dimensional spacetime.
The properties of these Kaluza-Klein particles are governed by the geometry of the extra dimension, and measurement
of these properties would yield information about the structure of space-time in the universe.
The parameter points in this work, which determine the Kaluza-Klein particle properties, were chosen to
be consistent with electroweak precision data, the $Z$-boson couplings to $b$-quarks, and flavor parameters
associated with kaon meson mixing.  The resulting signatures in several quark and lepton flavor observables
were examined, where it was found that such measurements would be powerful in distinguishing models
of warped extra dimensions; an example employing rare kaon decays and CLFV in muon transitions 
is presented in Fig.~\ref{fig:RS}.  The LHC signatures
for these benchmark models were also examined and correlated with the flavor observables.

The observation of non-zero neutrino masses points to the possibility of a new mechanism beyond the SM which
gives mass to the neutrinos, adding further structure over the SM Higgs mechanism.
A leading paradigm for
generating small neutrino masses is the so-called seesaw mechanism,
which 
posits the existence of a heavy right-handed
Majorana neutrino and leads to tiny masses through mixing.
The classic signal for both light and heavy Majorana neutrinos is the
presence of 
neutrinoless double-beta decay.
However this reaction may not be able to discern the full heavy-light
neutrino 
mixing effects, depending on the
exact value of the parameters.  It is possible to measure such mixing
effects at
 the LHC, if the seesaw scale
is near a TeV.  This synergy between the Intensity and Energy Frontiers shows that 
direct production of the right-handed neutrino and gauge boson at
the LHC was compared to the sensitivity in flavor-violating muon and $B$-meson decays,
as well in $0\nu\beta\beta$ \cite{Dev:2013vba}.  Probing this mixing at the LHC and ILC \cite{EFpaper} 
can provide crucial information
about the detailed nature of the seesaw mechanism, adding to the
knowledge 
gained from Intensity Frontier
experiments.

%% file: conclusions.tex
Here, we briefly summarize the main message from each working group.

\noindent{\bf The neutrino message:}
The past two decades have brought tremendous new information
about neutrinos, and we are now poised to build on our success. 
The experimental path forward is clear.  Long-baseline beam experiments will be major
players in fully testing the three-flavor paradigm.  There are multiple approaches
for determining the neutrino mass hierarchy; for leptonic CP violation, intense beams
will be required.  
While the three-flavor paradigm describes most of the data, wide gaps that could
accommodate new phenomena remain due to the absence of precise and detailed
information.  Such new physics may be revealed by  precision measurements, or by
experiments that pursue some of the existing intriguing anomalies. An array of
neutrinoless double-beta decay search experiments with different isotopes will
address the Majorana-vs-Dirac nature of the neutrino, and the question of the
absolute neutrino mass scale will be answered by kinematic experiments and precision
cosmology.

There is a large diversity of physics topics within the neutrino sector, and the
interplay between neutrino physics and other fields is rich.   Neutrinos have provided and
will continue to provide important information on structure formation in the early
universe, Earth, solar and supernova physics, nuclear properties, and rare decays of
charged leptons and hadrons.  Conversely, information regarding neutrino properties
and the origin of neutrino masses  is expected from the Energy and Cosmic Frontiers,
and from other areas of Intensity Frontier research, as well as nuclear physics.
In other words, neutrino physics sits at the nexus of the worldwide effort across the
frontiers of particle physics.

\noindent{\bf The baryon number violation message:}
Baryon number violation probes fundamental symmetries at very high energy scales. Estimates predict
fewer background events for a LArTPC, and the 34 kiloton LBNE and 560 kiloton
Hyper-Kamiokande, placed underground, have comparable sensitivity to proton decay modes.  A factor of 5 to 10 increase in
reach for the proton lifetime is expected, resulting in bounds on the lifetime of order $10^{34}$ to $10^{35}$ years.
This level matches that predicted by favored supersymmetric grand unified theories.  In addition, a new experimental
search for neutron-antineutron oscillations at Project X could improve existing limits by up to 4 orders of magnitude.
The observation of baryon number violation would have a profound impact on society as a whole due to the knowledge that 
the universe would have a finite lifetime.

\noindent{\bf The charged leptons message:}
The charged lepton experimental program offers significant discovery opportunities in this decade's experiments and
in even more sensitive experiments possible with future facilities such as Project X.  The experimental program
consists of a large and diverse set of opportunities and includes multi-purpose experiments that utilize the
large tau production rates at high-luminosity $B$ factories, as well as highly optimized single-purpose
experiments that explore muon transitions.  In particular E989 is under construction and is expected to settle
the issue of whether the value for the anomalous magnetic moment of the muon agrees with the SM prediction, or
provides a clear signal for the existence of new physics. Mu2e at Fermilab is expected to also come on-line during
this decade and provide an increase in sensitivity of several orders of magnitude 
to the direct conversion of a muon to an electron in the field of a nucleus.

Extremely sensitive searches for
rare transitions of muons and tau leptons, together with precision measurements of their properties, will either elucidate the 
scale and dynamics of flavor generation, or could limit the scale of new physics contributions to flavor generation to well above $10^4$ TeV.  Any
indication of charged lepton number violation would be an indisputable discovery of new physics.  Precision
measurements of lepton flavor-conserving processes can be used to verify predictions of the SM and look for signs
of new physics.

\noindent{\bf The quark flavor message:}
Quark flavor physics is an essential element in the international particle physics program.
Experiments that study the properties of highly suppressed decays of strange, charm, and bottom
quarks have the potential to observe signatures of new physics at mass scales
well beyond those directly accessible by current or foreseeable accelerators.
The importance of quark flavor physics is recognized in Europe and Asia,
as demonstrated by the commitments to LHCb, NA62, KLOE-2, and PANDA in Europe,
and to Belle~II, BESIII, KOTO, and TREK in Asia.  

In order for the U.S.\ particle physics program to have the breadth to assure
meaningful participation in future discoveries, significant U.S.\ contributions
to offshore quark-flavor experiments is important, and continued support
for U.S.\ groups in these efforts is a sound investment. 
In particular,  U.S.\
contributions to LHCb and Belle~II should be encouraged because of the richness
of the physics menus of these experiments and their reach for new physics.

Existing facilities at Fermilab are capable of mounting world-leading rare
kaon decay  experiments in this decade at modest incremental cost to running the
Fermilab neutrino program.   The proposed ORKA experiment, to measure the rare
decay $K^+ \to \pi^+ \nu \overline{\nu}$, with high precision, provides such an
opportunity to probe new physics with unmatched sensitivity.  
This is a compelling opportunity that should be exploited.
In the longer term, Project X can become the dominant facility in the
world for rare kaon decay experiments.  Its potential to provide ultra-high
intensity kaon beams with tunable time structure  is unprecedented.  While the
physics case for Project X is much broader than its capabilities  for kaon
experiments,  the power of a Project X kaon program is a strong argument in its
favor.

Collaboration between theory and experiment 
has always led to unexpected progress in
quark flavor physics, 
and this is expected to continue.  Therefore, stable support for
theorists working in this area is essential.   Lattice QCD plays a crucial role,
and support for the computing facilities needed for LQCD progress should be
maintained.

\noindent{\bf The nucleons, nuclei, and atoms message:}
Electric dipole moments of the electron, neutron and nuclei provide stringent tests for new sources of CP violation.
Present limits already place tight constraints on physics beyond the SM such as Supersymmetry.  A worldwide effort
is underway with several new experiments aiming to improve the EDM search reach by several orders of magnitude.  This
would reach the level of sensitivity of predictions from theories that attempt to explain the matter-antimatter
asymmetry of the universe.

\noindent{\bf The new light, weakly-coupled particles message:}
Dark sectors, and the new light weakly-coupled particles they may contain, are a well-motivated possibility 
for new physics. 
Axions, invented to solve the strong CP problem, are a perfect dark matter candidate. Dark photons and other 
hidden sector particles are equally compelling dark matter candidates. They could resolve outstanding 
puzzles in particle and astro-particle physics, and could explain dark matter interactions with the SM sector as well. 
Exploring dark sectors can proceed with existing facilities and comparatively modest experiments. 
A rich, diverse, and low-cost experimental program has been identified that has the potential for one 
or more game-changing discoveries. The U.S. high energy physics community should vigorously pursue these experiments.

\noindent{\bf The Intensity Frontier message:}
The working group reports and this summary show very clearly that the Intensity Frontier physics program is diverse
and rich in opportunities for paradigm-changing discovery.  Such an extensive and multi-pronged program is
necessary to address the unresolved fundamental questions about nature.  We are at a stage in our understanding where the SM
provides sharp predictive capabilities; however, we know it is not the final theory and there is more to learn about
the universe.  The knowledge we seek cannot be gained by a single experiment, or on a single frontier, but
rather from the combination of results from many distinct approaches working together in concert.

We hope this report provides valuable input into strategic decision-making processes in the U.S. and elsewhere,
and contributes to establishing a worldwide Intensity Frontier program in which the U.S. plays a leading
role.  This study confirmed that the proposed experiments will position the U.S. as a global
center in Intensity Frontier science.